\documentclass[sigconf]{acmart}

\AtBeginDocument{%
  \providecommand\BibTeX{{%
    \normalfont B\kern-0.5em{\scshape i\kern-0.25em b}\kern-0.8em\TeX}}}

\copyrightyear{2023}
\acmYear{2023}
\setcopyright{rightsretained}
\acmConference[CHI '23]{Proceedings of the 2023 CHI Conference on Human Factors in Computing Systems}{April 23--28, 2023}{Hamburg, Germany}
\acmBooktitle{Proceedings of the 2023 CHI Conference on Human Factors in Computing Systems (CHI '23), April 23--28, 2023, Hamburg, Germany}\acmDOI{10.1145/3544548.3580703}
\acmISBN{978-1-4503-9421-5/23/04}

\usepackage{xcolor,colortbl}
\usepackage{float}
\usepackage{subcaption}
\usepackage{multirow}
\definecolor{Gray}{gray}{0.85}
\definecolor{lightblue}{rgb}{0.68, 0.85, 0.9}

\newcolumntype{L}{>{\centering\arraybackslash}m{2cm}}
\newcolumntype{K}{>{\centering\arraybackslash}m{6cm}}
\newcolumntype{M}{>{\centering\arraybackslash}m{4cm}}

\begin{document}

\title[SwitchTube: An Adaptable Commitment Interface for Digital Wellbeing]{SwitchTube: A Proof-of-Concept System Introducing ``Adaptable Commitment Interfaces'' as a Tool for Digital Wellbeing}

\author{Kai Lukoff}
\orcid{0000-0001-5069-6817}
\affiliation{%
  \institution{Santa Clara University}
   \city{Santa Clara}
   \state{CA}
   \country{USA}
}
\email{klukoff@scu.edu}

\author{Ulrik Lyngs}
\orcid{0000-0001-9960-1436}
\affiliation{%
  \institution{University of Oxford}
  \city{Oxford}
  \country{UK}
}
\email{ulrik.lyngs@cs.ox.ac.uk}

\author{Karina Shirokova}
\orcid{0000-0001-7796-7509}
\affiliation{%
  \institution{University of Washington}
   \city{Seattle}
   \state{WA}
   \country{USA}
}
\email{karina27@uw.edu}

\author{Raveena Rao}
\orcid{0000-0002-2545-5066}
\affiliation{%
  \institution{University of Washington}
   \city{Seattle}
   \state{WA}
   \country{USA}
}
\email{rraveena@uw.edu}

\author{Larry Tian}
\orcid{0000-0003-4215-1635}
\affiliation{%
  \institution{University of Washington}
   \city{Seattle}
   \state{WA}
  \country{USA}
}
\email{ganlin@uw.edu}

\author{Himanshu Zade}
\orcid{0000-0003-1755-7780}
\affiliation{%
  \institution{University of Washington}
   \city{Seattle}
   \state{WA}
   \country{USA}
}
\email{himanz@uw.edu}

\author{Sean A. Munson}
\orcid{0000-0002-0472-6138}
\affiliation{%
  \institution{University of Washington}
   \city{Seattle}
   \state{WA}
   \country{USA}
}
\email{smunson@uw.edu}

\author{Alexis Hiniker}
\orcid{0000-0003-1607-0778}
\affiliation{%
  \institution{University of Washington}
   \city{Seattle}
   \state{WA}
   \country{USA}
}
\email{alexisr@uw.edu}

\renewcommand{\shortauthors}{Lukoff et al.}

\begin{abstract}
YouTube has many features, such as homepage recommendations, that encourage users to explore its vast library of videos. However, when users visit YouTube with a specific intention, e.g., learning how to program in Python, these features to encourage exploration are often distracting. Prior work has innovated `commitment interfaces' that restrict social media but finds that they often indiscriminately block needed content. In this paper, we describe the design, development, and evaluation of an `adaptable commitment interface,' the SwitchTube mobile app, in which users can toggle between two interfaces when watching YouTube videos: Focus Mode (search-first) and Explore Mode (recommendations-first). In a three-week field deployment with 46 US participants, we evaluate how the ability to switch between interfaces affects user experience, finding that it provides users with a greater sense of agency, satisfaction, and goal alignment. We conclude with design implications for how adaptable commitment interfaces can support digital wellbeing.
\end{abstract}



\begin{teaserfigure}
    \includegraphics[width=\textwidth]{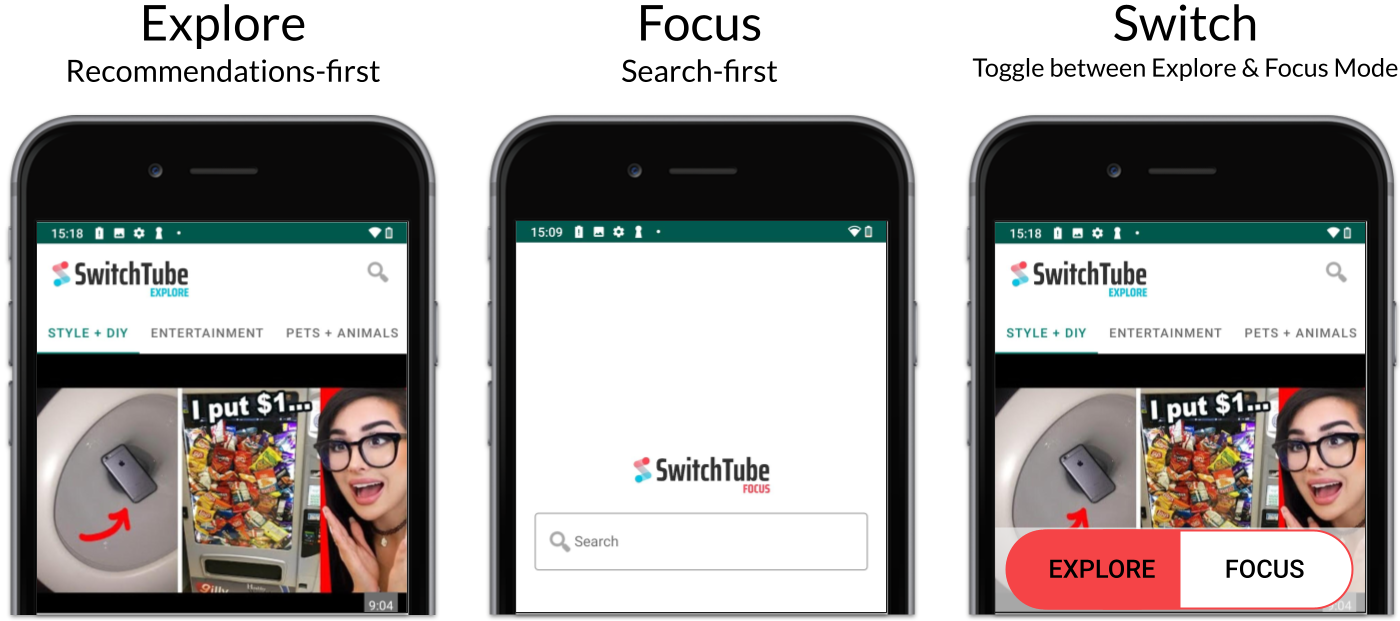}
   \caption{The three different versions of the SwitchTube study app are designed to support different levels of sense of agency for the user. The Explore Version (low sense of agency) has a recommendations-first homepage whereas the Focus Version (medium) has a search-first homepage. The Switch Version (high) is an ``adaptable commitment interface'' that lets users toggle between Explore and Focus Mode.}
   \Description{Three different versions of the SwitchTube app, which manipulates the level of control users have when watching YouTube videos.}
   \label{3versions_switchtube}
   \Description{Image displaying three unique home page interfaces for the SwitchTube application. The Explore version prioritizes recommendations, so the homepage contains a list of suggested videos. The Focus version prioritizes search, so the homepage only contains a search box. The Switch version includes both Explore and Focus settings, and users can toggle between Explore and Focus using the toggle.}
 \end{teaserfigure}

\maketitle

\section{Introduction}

Temptations, i.e., desires that conflict with long-term goals, are common in everyday life. When people attempt to resist everyday temptations, they report a failure rate of about 30\% for smoking, 30\% for drinking alcohol, and 35\% for shopping. However, the temptation with the highest self-control failure rate of all is media use, e.g., watching TV or checking social media, which people report failing to resist 76\% of the time according to the results of a day-reconstruction study \cite{Delaney2017-ld}. Media use is again associated with a far higher failure rate than other temptations in a similar study using the experience sampling method \cite{Hofmann2012-uv}.\\

Users often blame themselves for a lack of self-control \cite{24_Lanette2018-yc}, yet the alluringness of media use is no accident. Social media is intentionally designed by companies who have a financial interest in capturing and monetizing user attention \cite{Wu2016-ke}. As Tristan Harris, who co-founded the advocacy group Time Well Spent, notes, \textit{“there’s a thousand people on the other side of the screen whose job is to break down whatever responsibility I can maintain”} \cite{Bosker2016-du}. Individually the costs of these short-term distractions may seem small, but collectively they undermine people’s ability to spend time in accordance with their values \cite{Williams2018-co,Zuboff2019-gs}.\\

YouTube, one of the most-watched video services in the world \cite{Lebow2022-ga}, is a prominent example of a service that is intentionally designed as a minefield of media use temptations. For example, YouTube has a rich selection of educational content, but when users go to watch it the recommender system often shows more entertaining videos that distract them away \cite{Klobas2019-cl}. And even when users do want entertainment, the system often sucks them into the YouTube `rabbit hole' of watching more than they intended \cite{Woolley2022-ho}. Specific features within YouTube—such as autoplay, related videos, and homepage recommendations—systematically induce this diminished user sense of agency, i.e., an individual’s experience of being the initiator of their actions in the world \cite{27_Lukoff2021-gm}. \\

In response, designers have innovated `commitment interfaces' in which the user sets a goal of limiting time on distracting apps or services and the interface holds them to it \cite{29_Lyngs2022-ug}. The `screen time tools' that now come pre-installed on all iOS and Android devices function this way. Unfortunately, this approach also indiscriminately blocks access to content and features within apps that users still want to access (e.g., educational tutorials on YouTube or the groups feature in Facebook) \cite{26_Lukoff2022-xs}. \\

Instead of forcing users into an all-or-nothing bargain, one alternative is to offer an \emph{adaptable commitment interface}, in which the user can toggle between different interfaces to meet their situational needs. However, a key question is whether such an adaptable interface provides a sufficient enforcement mechanism to hold the user to their commitment, or whether the ability to easily switch to a version with temptations will undermine the user’s digital self-control \cite{30_Lyngs2019-xr}. \\

In this research, we first surveyed a general population of 606 YouTube users from the US to understand their goals and motivations for changing their use. About half of participants were actively trying to change their YouTube use or planned to do so within the next month: about one-third of participant goals  were about reducing quantity of use, while about half were about shifting the \textit{quality} of videos consumed (e.g., watch fewer comedy videos, but more educational ones). Survey participants who wanted to change their YouTube use were further invited to participate in a field experiment with the SwitchTube mobile app. \\

We designed and developed three versions of SwitchTube: Explore (recommendations-first), Focus (search-first), and the Switch Version, an adaptable commitment interface, in which users can toggle between the Explore and Focus interfaces (Figure \ref{3versions_switchtube}). In a three-week field experiment, 46 participants used each of the these three versions for one week. Triangulating log data, experience sampling responses, and exit interviews, we identified four key use cases in which toggling between interfaces helped users. For example, users valued being able to start with a focused search and then explore from there, rather than \textit{immediately} landing on a homepage full of juicy recommendations unconnected to their intention. The data also revealed that the Switch Version generally provided users with the `best of both worlds' in terms of user experience: the ability to focus on goals, \textit{without} severely restricting their agency to explore new and satisfying content. \\

\begin{table*}
    \centering
    \setlength{\tabcolsep}{10pt}
    \renewcommand{\arraystretch}{1}
    \caption{A taxonomy of the effects of media consumption. Adapted from \cite{17_Gui2009-pc}}
    \small
    \begin{tabular}{|c c c|}
    \hline
    \rowcolor{Gray}
    & \textbf{Satisfaction} & \textbf{Wellbeing} \\
    \rowcolor{Gray}
    & (short-term) & (long-term) \\
    \hline
    \cellcolor{Gray}\textbf{Quantity} of TV & \multirow{2}{*}{Time choice effects, e.g., regret for lost time} & \multirow{2}{*}{Lifestyle effects, e.g., loss of sleep} \\
    \cellcolor{Gray}consumption & & \\
    \hline
    \cellcolor{Gray}\textbf{Quality} of TV & \multirow{2}{*}{Content choice effects, e.g., dissatisfaction with an episode} & \multirow{2}{*}{Cultivation effects, e.g., fostering increased materialism} \\
    \cellcolor{Gray}consumption & & \\
    \hline
    \end{tabular}
    \label{tab1:effectofmediaconsumption}
\end{table*}

This paper contributes (1) the results of large survey that helps to understand how social media users want to shift their use, not just reduce it; (2) the design, development, and evaluation of a proof-of-concept system demonstrating that adaptable commitment interfaces can improve both short-term user satisfaction \textit{and} long-term user goal alignment in the wild; and (3) implications for how to design adaptable commitment interfaces to support user sense of agency.

\section{Background and Motivation}

We briefly review related work on problematic YouTube use and then consider the opportunity and challenges for adaptable commitment interfaces as a new design approach.

\subsection{YouTube as a Site of Problematic Technology Use}

YouTube is a source of entertainment and education, but for many also of distractions that conflict with their preferences and goals \cite{Balakrishnan2017-cb} and lead to a sense of regret \cite{Cho2021-ng}. The authors of a case study of problematic YouTube use write, \textit{``[Adam] reflects that he was initially attracted to YouTube for its educational value, but `instead, I focused on music videos, funny videos and for entertainment purposes only''} \cite[p.273]{Klobas2019-cl}. In line with this quote, past work suggests that users have concerns over not only the quantity of content they consume, but also the quality (e.g., the ratio of entertainment to educational videos they watch) \cite{50_noauthor_undated_fm,Alter2017-wy} \\

In a study of television consumption \cite{17_Gui2009-pc}, Gui and Stanca introduce a taxonomy for understanding quantity and quality in both the short-term and long-term as dimensions of problematic media use (Table \ref{tab1:effectofmediaconsumption}). Current `screen time' tools (e.g., Apple's Screen Time) tend to address the quadrant for time choice effects, by limiting time spent on phones or in certain apps. Yet they largely do not alter the content that users encounter within apps \cite{27_Lukoff2021-gm}, although the aforementioned work on YouTube suggests that the quality-related quadrants are also important for digital wellbeing designers to consider. This includes both short-term `content choice' effects such as the way that watching a video makes one feel afterwards and 
long-term `cultivation effects' such as how the content of television shapes one's values and preferences over time. As a starting point, designers might benefit from a better understanding of \textit{how} users want to change the quality of the content they consume on YouTube so that they can create tools that help users focus on the content that supports their short-term satisfaction \textit{and} long-term well-being.

\subsection{Commitment Interfaces for Digital Self-Control}

Recognizing that many users struggle to manage their social media use, design practitioners and researchers have created hundreds of digital self-control tools to support them \cite{30_Lyngs2019-xr}. These tools leverage many different behavior change strategies, including raising awareness of (over)use \cite{Collins2014-tt,Kim2016-fy}, applying social pressure and accountability \cite{Ko2015-yc,Ko2016-xy}, and affirming the user’s goals and values \cite{31_Lyngs2020-jp,18_Hiniker2016-cs}. \\

\begin{figure}[H]
    \centering
    \includegraphics[scale=0.4]{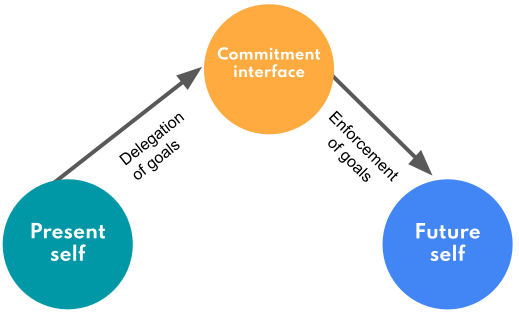}
    \caption{The present self does not trust the future self to stick to their goal, so they delegate their goal to a digital self-control tool which acts as a commitment interface. The tool enforces the goals upon the future self. But if the enforcement is too easy to override, will the commitment still be effective? Reprinted from \cite{26_Lukoff2022-xs}}
    \label{fig2:presentselfFutureself}
    \Description{A diagram depicting the connection between the present self and the commitment device, and the commitment device and the future self. The relationship between the present self and the commitment device is delegation of goals, while the relationship between the commitment device and the future self is enforcement of goals.}
\end{figure}

But perhaps the most common type of tool is what can be called a ‘commitment interface,’ in which the user commits to a goal of changing their use of a particular feature or app and the tool holds them to it \cite{Bryan2010-wt,Plaut2014-sw}. That is, the present self engages in reflective `System 2' thinking and anticipates that their future self, when engaged in automatic `System 1' thinking, will be unable to stick with their goal \cite{Kahneman2011-gt}. So the present self delegates their behavioral goals to a digital self-control tool (Figure \ref{fig2:presentselfFutureself}). The digital self-control tool then acts as a commitment device: it enforces the wishes of the present self upon the future self. For example, a user might resolve to limit their Facebook use to 30 minutes a day and install a browser extension that locks them out when time is up. \\

In a review of 367 digital self-control tools, Lyngs et al. found that about 50\% of tools on the Chrome browser store minimized distracting features (e.g., hiding Facebook's newsfeed) and about 60\% of tools on Google Play imposed blocks on apps \cite{30_Lyngs2019-xr}. 
Among tools created as part of academic research, LocknType used a lockout task, forcing the user to type an arbitrary string of digits in order to access distracting apps \cite{Kim2019-cd}. TypeOut combined a lockout task with a self-affirmation task, finding that the two interventions had an additive effect in reducing phone usage \cite{47_Xu2022-qw}. GoalKeeper imposed more significant restrictions on use, setting a time goal and keeping users locked out of their phone at successively longer intervals when it was exceeded \cite{22_Kim2019-pq}. HabitLab tested interventions with different levels of challenge, finding that users tend to start with the difficult ones, but gradually move to easier ones, although they never give up hope that they will return to the more difficult ones \cite{Kovacs2021-sz}. \\

\begin{figure}[H]
\label{fig3:enforcement_mechanisms}
\centering
\begin{subfigure}[H]{0.4\linewidth}
\caption{\hspace{0.4cm}\centering{\textbf{Weak enforcement}\newline
Easily tolerated by users, but goals are also easily and frequently circumvented}
}
\label{fig3:PanelA}
\includegraphics[width=\linewidth]{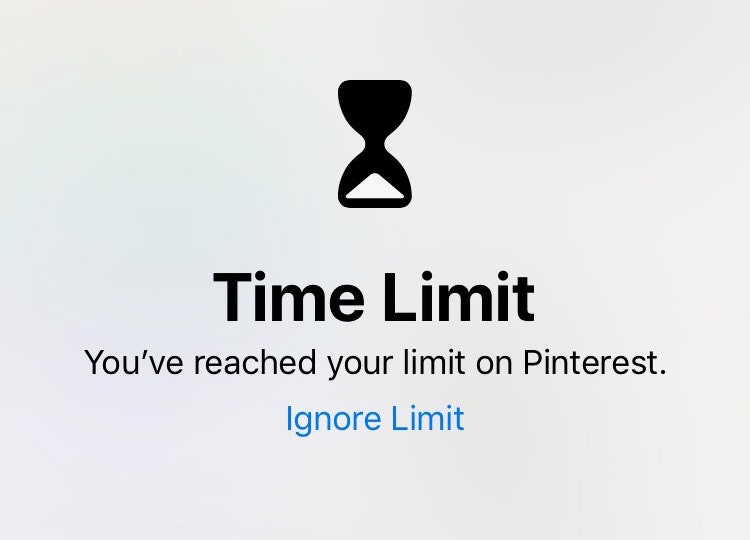}
\caption*{\centering Apple’s Screen Time lets users easiy ignore their time limit in apps}
\end{subfigure}
\hspace{0.2cm}
\begin{subfigure}[H]{0.5\linewidth}
\caption{\hspace{0.5cm}\centering{\textbf{Strong enforcement}\newline Helps users reach their goals, but often triggers frustration that leads to abandonment of the tool}
}
\label{fig3:panelB}
\includegraphics[width=\linewidth]{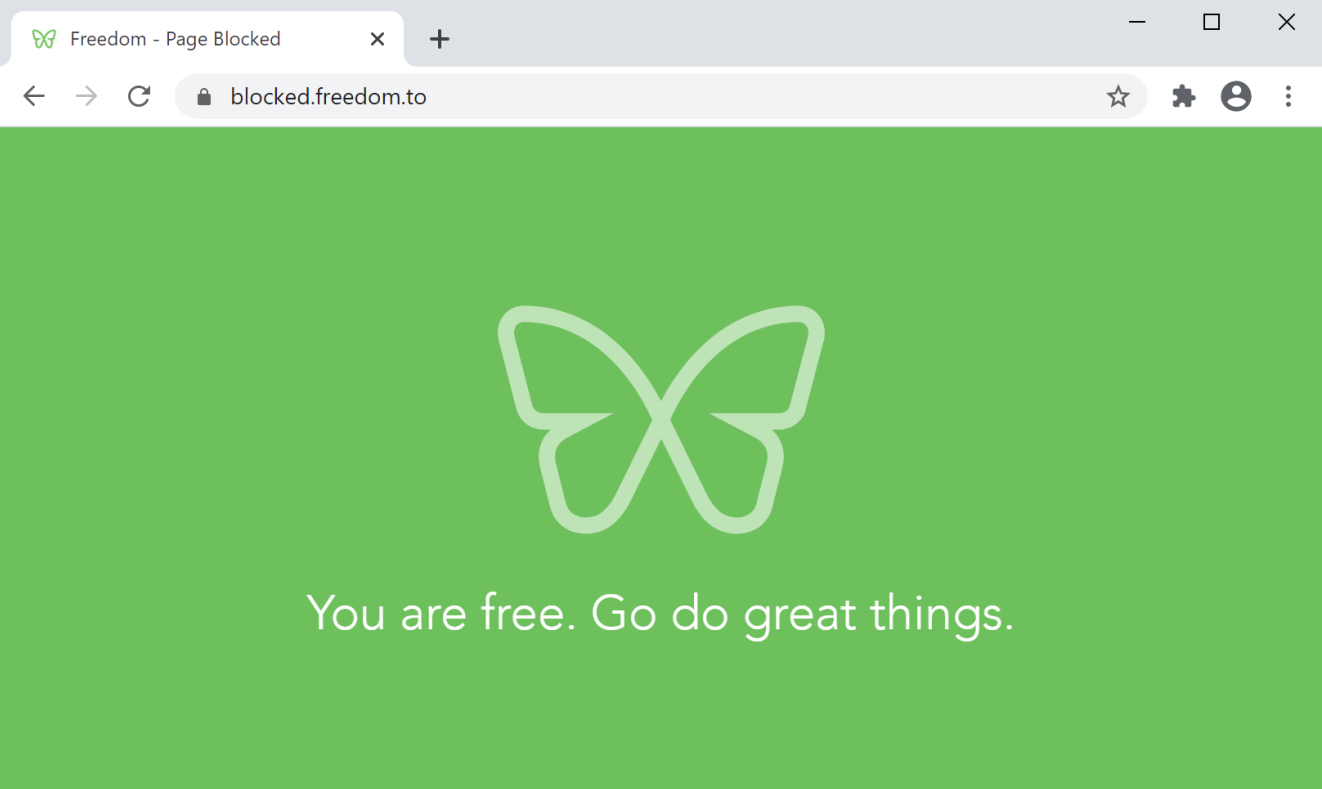}
\caption*{\centering The desktop software Freedom blocks websites without an override option}
\end{subfigure}%
\caption{Enforcement mechanisms in commitment interfaces}
 \Description{One image on the left demonstrates how Apple's screen time allowed users to easily bypass their time limit in apps such as Pinterest, while the image on the right demonstrates how the desktop software freedom blocks websites without an override option using a green interface with only a butterfly image and the text ``You are free. Do what matters.”}
\end{figure}

A universal challenge for `commitment interfaces’ is finding the right level of strictness. Prior work has found that digital self-control tools are generally more effective at curbing screen time when the severity of enforcement or the degree of friction of the tool is stronger \cite{10_Cox2016-ts,22_Kim2019-pq}, an intuition that users themselves hold \cite{42_Tran2019-gg,47_Xu2022-qw}. Too weak and it might be too easy for the user to circumvent their original goal, for example by clicking “Ignore Limit” on a warning that time is up (Figure \ref{fig3:PanelA}). But too strong and it might trigger frustration and lead them to abandon the tool completely, as in software that blocks a website with no override option (Figure \ref{fig3:panelB}). This motivates a search for a ``Goldilocks'' level of enforcement that strikes a good balance: providing the right amount of goal support without leading to psychological reactance against the tool \cite{22_Kim2019-pq,29_Lyngs2022-ug}. We now turn to adaptable commitment interfaces as a new design approach for navigating this challenge. 

\subsection{Adaptable Commitment Interfaces for Digital Wellbeing}
\begin{figure}[ht]
\centering
\begin{subfigure}[H]{0.4\linewidth}
\frame{\includegraphics[width=\linewidth]{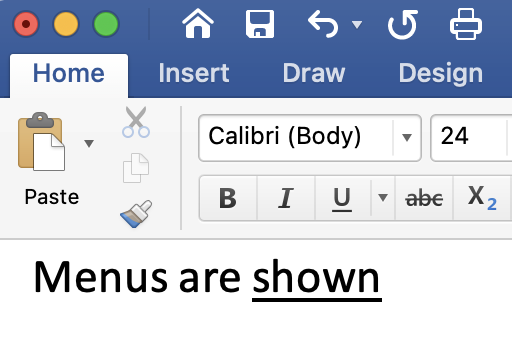}}
\caption{Focus Mode OFF}
\end{subfigure}
\hspace{0.3cm}
\begin{subfigure}[H]{0.4\linewidth}
\frame{\includegraphics[width=\linewidth]{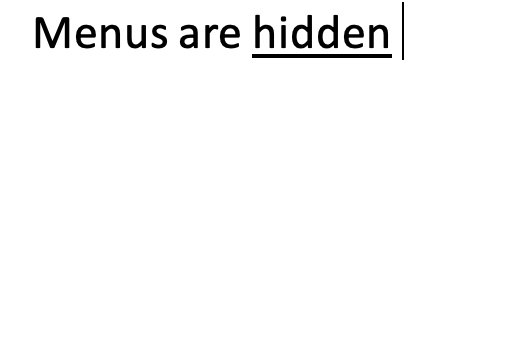}}
\caption{Focus Mode ON}
\end{subfigure}%
\caption{The Microsoft Word desktop app enables the user to manually turn on “Focus Mode”, which hides the menus and toolbar.}
\label{fig1:focusmodes}
 \Description{Two images of the Microsoft Word desktop application with the focus mode enabled and disabled. When the focus mode is not active, the interface features a toolbar with numerous options. When the focus mode is active, the app interface displays only a blank page containing the text the user has input.}
\end{figure}

Adaptable interfaces are interfaces that the user can change to fit their situational needs \cite{Gulla2015-mn}. For example, in the Microsoft Word desktop app the user can manually select ``Focus Mode'' to hide the toolbar and menus and write without distraction (Figure \ref{fig1:focusmodes}). As another example, the Google Maps mobile app automatically turns on `Dark Mode' after sunset to make it easier to see driving directions at night. But the concept of ‘adaptable interfaces’ might also be applied beyond just productivity tools. \\

Many social media users also have a need for different levels of control at different times. For instance, past work has found that when YouTube users have a specific intention in mind, they tend to prefer a search-first interface. But when users do not have a specific intention in mind, they prefer to turn control over to YouTube with a recommendations-first interface \cite{27_Lukoff2021-gm}. One potential solution then is to let the user change the interface to support the level of control that they want in that particular situation. This might be for a single feature, e.g., YouTube currently offers an on/off toggle for autoplay. Or it might be for a bundle of multiple features, e.g., switching between an Explore Mode and a Focus Mode. \\ %

When an adaptable interface supports the user in exercising different levels of control over temptations, we might call it an adaptable commitment interface (ACI). Our review suggests competing hypotheses for this type of interface. On the one hand, an ACI might provide the best of both worlds: an interface that lets users determine which interface best suits their needs for that situation. On the other hand, if the user can too easily switch between interfaces, an ACI might undermine the enforcement mechanism that the present self uses to hold the future self to their goals. This is a unique challenge that does not apply to most cases of adaptable interfaces (e.g., Dark Mode on Google Maps, where switching to Light Mode is not a temptation). However, in the case of digital self-control tools where there is a temporal conflict in user goals, it remains an open question whether an adaptable interface still functions as an effective commitment device.

\subsection{Research Questions}
We address three research questions regarding YouTube use and adaptable commitment interfaces:
\begin{itemize}
    \item \textbf{RQ1:} \textit{How, if at all, do U.S. YouTube users want to change the quantity and quality of their video consumption?} We draw upon the stages of change model of motivation \cite{37_Prochaska2013-nh} to understand how strongly and in what direction people want to change their YouTube use.
    \item \textbf{RQ2:} \textit{When and why do people switch in an adaptable commitment interface?} In the Switch Version of the app, users can toggle between Focus and Explore Mode, which are designed to support different levels of sense of agency. Since users have the choice to switch at any time, will they switch in ways that undermine their intentions? Or will they switch in ways that meet their situational needs?
    \item \textbf{RQ3:} \textit{How does an adaptable commitment interface influence user experience?} In particular, how do the three versions influence measures of short-term satisfaction and long-term goal alignment?
\end{itemize}

\noindent We first report on a survey of the change motivations that YouTube users hold in the first place (RQ1), before turning to the design and deployment of SwitchTube, an adaptable commitment interface designed to answer RQ2 and RQ3.

\section{Survey: Stages of Change for YouTube Users}
In behavior change, the stages of change model (also known as the transtheoretical model) is often used to assess an individual’s readiness to make a change \cite{37_Prochaska2013-nh}, ranging from pre-contemplation (no intention to change behavior) to action (actively trying to modify behaviors). We draw upon this model to answer \textbf{RQ1}: \textit{How, if at all, do U.S. YouTube users want to change the quantity and quality of their video consumption?} We conduct a large-scale survey to assess what percentage of users are motivated to change and the nature of their digital wellbeing goals. This information can inform the design of alternative versions of social media and YouTube, as we do with the SwitchTube app in this work.

\subsection{Methods}
\textbf{Recruitment:} To obtain a general sample of users of the YouTube mobile app, we recruited participants from Prolific, a platform that specializes in crowdsourcing participants for research studies. Participants were invited to a survey about their \emph{``YouTube watching habits and preferences.''} We sought to recruit a general population of YouTube users, so intentionally did \textit{not} advertise for participants who wanted to change or reduce their YouTube use. Participants were required to meet four inclusion criteria: live in the United States, be fluent in English, use YouTube at least once per month, and use an Android mobile phone.\\

\noindent \textbf{Demographics:} A total of 606 participants met the inclusion criteria and completed the survey (see demographics in Table \ref{tab2:606Demographics}). We excluded responses from an additional 90 participants who started but did not complete the survey. We happened to oversample young people relative to general U.S. population \cite{43_United_States_Census_Bureau_undated-re}.

\begin{table}[H]
    \centering
    \caption{Demographics of the 606 survey participants}
    \begin{tabular}{|>{\columncolor[gray]{0.85}} l|p{5cm}|}
    \hline
         Gender identity & Man (53\%), Woman (47\%), Non-binary (0\%), Prefer not to say (0\%)  \\
         Age range & 18-24 (19\%), 25-34 (39\%), 35-44 (25\%), 45-54 (9\%), 55+ (8\%) \\
         Education & High school (34\%), Associate degree (22\%), Bachelor’s degree (46\%), Advanced degree (11\%) \\
         Race/ethnicity & White (64\%), Black (12\%), Asian (9\%), Hispanic (6\%), 2 or more races (8\%), other (1\%)\\
         \hline
    \end{tabular}
    \label{tab2:606Demographics}
\end{table}

\noindent \textbf{YouTube Use:} Survey participants spent a median of 90 minutes per day (interquartile range: 40-180) on YouTube across all devices in the week prior to the survey (self-estimated\footnote{Self-estimates of time spent on social media are only moderately correlated with actual usage \cite{14_Ernala2020-pf}. We considered asking participants to report the weekly Time Watched stats presented by the YouTube app, however the help page for YouTube currently states, \emph{``Due to a known issue, time watched on computers is reported incorrectly''} without further explanation, rendering this data source highly questionable.}). Of this time, participants estimated they spent a median of 52\% (interquartile range: 20-80\%) in the mobile app.\footnote{For comparison, the YouTube press page states that mobile accounts for over 70\% of watch time \cite{40_Solsman2018-zi}.} Upon multiplying time on all devices with the percentage spent on mobile, for each participant, we found that participants spent an average of 34 minutes per day in the YouTube mobile app. This is below the average for all YouTube users: in 2017, YouTube shared that signed-in users spend an average of more than 60 minutes per day in the mobile app \cite{32_Matney2017-kp}.

\subsection{Procedure}
In our online survey, participants first answered background questions about their demographics, technology use, and YouTube use specifically. \\

To investigate \textbf{RQ1}, we adapted a questionnaire from prior behavior change research \cite{46_Wright2011-kh} to assess participants’ stage of change with regard to their YouTube use (Table \ref{tab3:tab_stagesofchange}). For example, our first question asked, \emph{``Are you currently trying to take more control over how you spend time on YouTube?''} If the participant answered \emph{``Yes''}, they were categorized into the Action Stage. If they answered \emph{``No''} or \emph{``Not sure''}, the survey asked further questions about their readiness to change. \\

To understand the nature of their intended change, we asked, \emph{``What, if anything, do you want to change about your YouTube use?''} and how important that change was for them (5-point scale; Not at all important - Extremely important). Participants were paid \$1.37 for a survey that took an estimated 8 minutes to complete, an incentive rate that exceeds the US minimum wage (\$7.25 per hour). This research was approved by the University of Washington IRB.

\subsection{Coding Reliability Thematic Analysis}
We conducted a coding reliability thematic analysis \cite{5_Braun2018-th} to identify commonalities in participant goals for change. Three authors read through 100 of the 606 responses together and discussed possible codes informed by Gui and Stanca's framework of the quantity and quality of media overuse \cite{17_Gui2009-pc}. {They created a shared codebook with mutually inclusive codes and subcodes. For example, the change response,  \textit{``I would like to watch more educational videos rather than a lot of the junky videos that pop up on my feed''} was coded as ``Increase type of content (code) $\rightarrow$ 
Educational or useful (subcode)'' and also as ``Decrease type of content (code) $\rightarrow$ Recommendations (subcode).'' Two authors then applied this codebook across a random sample of 50 further responses. The interrater reliability for our codes and subcodes ranged from Cohen's Kappa of 0.66 (substantial agreement) to 1 (perfect agreement)} \cite{23_Landis1977-xl}. These two authors then used the codebook to each code half of the remaining responses.
\begin{table*}
    \centering
    \caption{The questions and responses used to categorize survey participants into stages of change with regard to their YouTube use.}
    \begin{tabular}{|p{10.5cm} | p{2.5cm} | p{2.5cm}|}
    \hline
    \rowcolor{Gray}
    \multicolumn{1}{|c}{\textbf{Question}} & \multicolumn{1}{|c}{\textbf{Stage of change}} & \multicolumn{1}{|c|}{\textbf{Participants (\%)}} \\
    \hline
    Are you currently trying to take more control over how you spend time on YouTube? & & \\
    \multicolumn{1}{|l|}{Yes $\xrightarrow{\makebox[9.5cm]{}}$ } & Action & \multicolumn{1}{c|}{40\%} \\
    No or Not sure >> (go to the next question) & & \\
    \hline
    Do you plan to [...] in the \textbf{next month}? & & \\
    Yes $\xrightarrow{\makebox[9.5cm]{}}$ & Preparation & \multicolumn{1}{c|}{8\% } \\
    No or Not sure >> (go to the next question) & & \\
    \hline
    Do you plan to [...] in the next \textbf{six months}? & & \\
    Yes $\xrightarrow{\makebox[9.5cm]{}}$ & Contemplation & \multicolumn{1}{c|}{4\% } \\
    No or Not sure >> (go to the next question) & & \\
    \hline    
    Do you \textbf{currently} feel in control of how you spend time on YouTube? & & \\
    Yes $\xrightarrow{\makebox[9.5cm]{}}$ & Maintenance  & \multicolumn{1}{c|}{45\% } \\
    No or Not sure >> (go to the next question) & Precontemplation & \multicolumn{1}{c|}{3\% } \\
    \hline
    \end{tabular}
    \label{tab3:tab_stagesofchange}
\end{table*}
\begin{table*}
    \centering
    \caption{Our coding of the changes to YouTube use that the 606 participants in our survey wanted to make.\protect\footnotemark}    
    \begin{tabular}{p{5.5cm}cp{5.5cm}c}
    \hline
    \rowcolor{Gray}
    \multicolumn{1}{c}{\textbf{Goal Code}} & \multicolumn{1}{c}{\textbf{\% of codes}} & \multicolumn{1}{c}{\textbf{Sub-code}} & \multicolumn{1}{c}{\textbf{\% of parent code}}\\
    \hline
    Increase quantity of time &	5\% & & \\
    Decrease quantity of time &	37\% &	During work &	8\% \\
     & & During nighttime &	7\% \\
     Increase type of content &	28\% &	Meaningful &	12\% \\
     & & Educational or useful &	34\% \\
     & & Entertaining, funny, interesting &	12\% \\
     & & New or diverse &	18\% \\
     & & Other &	25\% \\
     Decrease type of content &	23\% &	Ads &	20\% \\
     & & Meaningingless &	27\% \\
     & & Entertaining, funny, interesting &	7\% \\
     & & Recommendations &	15\% \\
     & & Other &	23\% \\
     Increase awareness of use &	2\% & & \\
     Other &	8\% & & \\
     \hline
    \end{tabular}
    \label{tab4:goalcodes}
\end{table*}

\subsection{Result and Analysis}
\textbf{Stages of change.} Based on participant survey responses (Table \ref{tab3:tab_stagesofchange}), we categorized participants into 5 different stages of change: 
\footnotetext{Subcodes that accounted for less than 5\% of the parent code are not displayed.}
\begin{enumerate}
    \item \textbf{Action} (40\%): actively trying to change
    \item \textbf{Preparation} (8\%): planning to change shortly
    \item \textbf{Contemplation} (4\%): wanting to change, but with no immediate plans
    \item \textbf{Precontemplation} (3\%): no intention of changing
    \item \textbf{Maintenance} (45\%): maintaining current behavior
\end{enumerate}

Responses followed a bimodal distribution, with the vast majority of participants either at the Action Stage (40\%) or the Maintenance Stage (45\%). That is, participants were predominantly either actively trying to make a change or already satisfied with their behavior. We missed the opportunity to ask participants in the maintenance stage whether they had previously made a change that they were now maintaining or whether they had simply been satisfied with their YouTube use all along. Overall, about half of this sample of general YouTube users were either actively trying or in preparation to make a change to take more control of their YouTube use.\\

The concept of ``lagging resistance'' documented for Facebook users \cite{1_Baumer2013-qp} suggests that many social media users might fall into the murky middle of wanting to change but not doing so just yet. However, our results find that YouTube users who wanted to change were trying to do so now or imminently. We speculate that this may be because YouTube users face fewer \textit{social} pressures,  whereas Facebook users often feel like they cannot reduce their use because friends, family, and colleagues expect them to be active and responsive on Facebook. The fact that about half of participants were in the action or preparation stage aligns with previous findings of an underlying discontent with social media use \cite{24_Lanette2018-yc}. \\

\textbf{Nature of Goals.} Of the 606 survey participants, 71\% shared goal(s) for change and 29\% did not want to change anything. In other words, even among the 45\% of participants who were in the maintenance stage and already felt in control of the time they spent on YouTube, there was still often a desire to change \textit{something} about their YouTube use.\\

\begin{figure*}
    \centering
    \includegraphics[width=\textwidth,height=11cm]{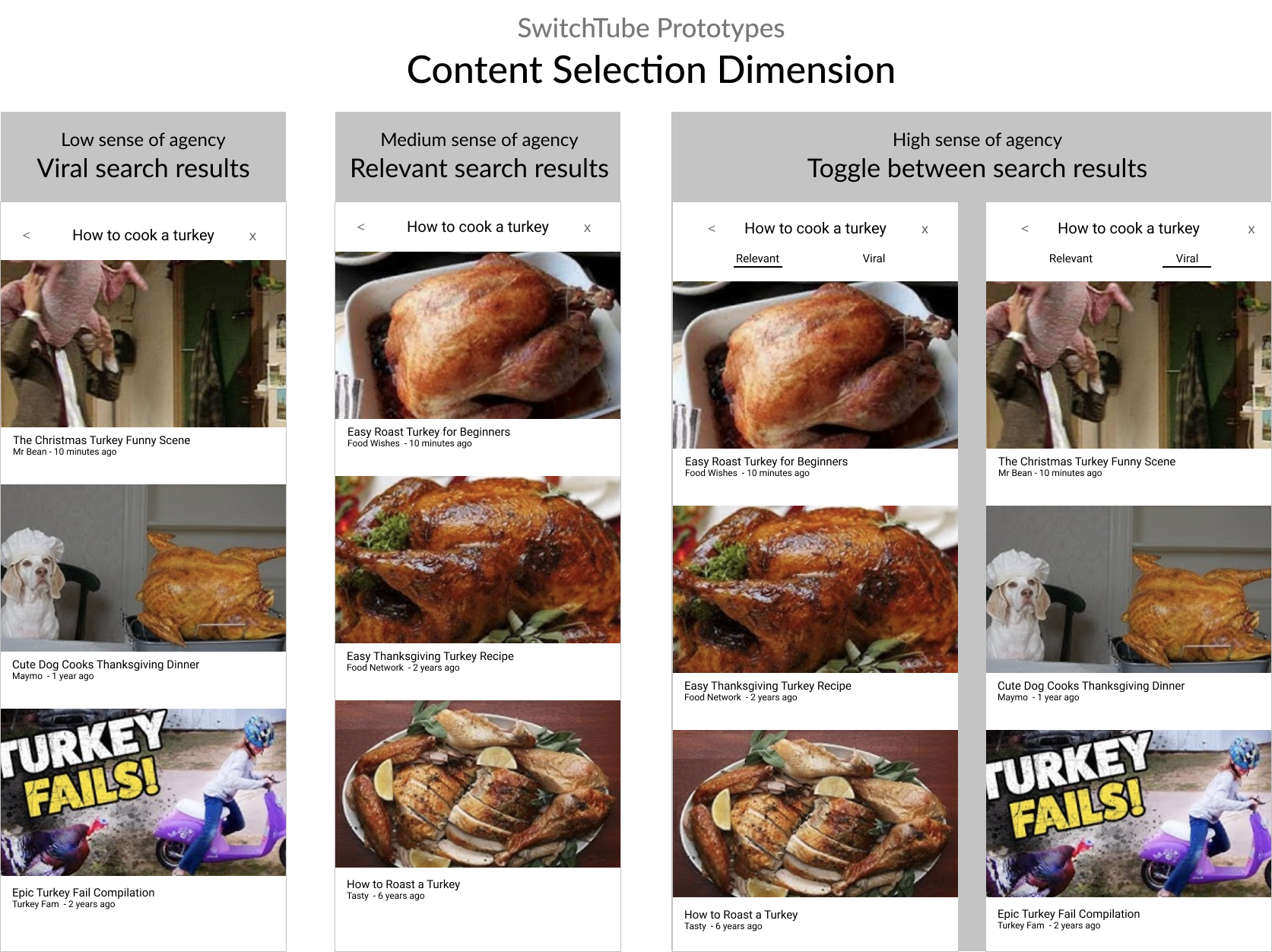}
    \caption{For the Content Selection dimension, we showed paper prototypes of three different versions of search results that varied entertainment level to support a low, medium, or high sense of agency. Here, when a user searches for “How to cook a turkey” they see either what we called viral results (e.g., “Cute Dog Cooks Thanksgiving Dinner”) or relevant results (e.g., “Easy Roast Turkey for Beginners”), and one of the versions gives the user to switch between the two types of results. Based on the user feedback we received on these prototypes, we hypothesized that the Switch Version might actually provide the greatest sense of agency.}
    \label{fig4:fig_prototypes}
    \Description{An image showing the difference between the viral search results (with content that would catch people’s attention) and the relevant search results (with different turkey cooking images), and how users see those two different search results with a toggle.}
\end{figure*}

Participants shared a total of 817 goals and wanted to change both the quantity and quality of time that they spent on YouTube (Table \ref{tab4:goalcodes}). In terms of quantity, participants overwhelmingly wanted to decrease the amount of time they spent on YouTube (37\% of goals) rather than increase it (5\%). Participants mentioned working hours and bedtime as particular times that they wanted to reduce use: \emph{``I tend to watch it a bit too much, especially as background noise while I work from home.''} In terms of quality, participants mentioned some content that they wanted to watch more of (28\% of goals) and other content that they wanted to watch less of (23\%). Participants wanted to watch more content that was educational (34\%), new (18\%), entertaining (12\%), and meaningful (12\%). Participants wanted to watch less content that was meaningless (27\%), ads (20\%), recommendations (15\%), and entertaining (7\%). Less common goals included increasing awareness of use (2\%) and a diverse category of other goals $(8 \%)$ such as creating videos, re-organizing saved videos, and reducing data usage.\\

Of note is that the screentime tools that come pre-installed on almost all smartphones (Apple Screen Time and Google's Digital Wellbeing) currently address only about 40\% of the change goals that participants had. Given their focus on limiting time spent and presenting usage stats, they might support the goals of decreasing overall quantity of time (37\% of goals) and increasing awareness (2\%). However, they do not have features that support participants to change the type of content they consume within apps such as YouTube. For these goals, participants have to turn to the features of YouTube itself, where previous research has found that users feel frustration over their inability to customize the quantity, content, and placement of recommendations that appear in the interface \cite{27_Lukoff2021-gm}.\\

\begin{table*}
    \centering
    \caption{Three Versions of SwitchTube}
    \begin{tabular}{p{1.5cm} p{2.5cm} p{3.5cm} p{3cm} p{4cm}}
         \hline
         \rowcolor{lightblue}
         \textbf{Version} &
         \textbf{Intended level of sense of agency}&
         \multicolumn{1}{c}{\textbf{Homepage}}&
         \multicolumn{1}{c}{\textbf{Search Results}}&
         \multicolumn{1}{c}{\textbf{Video Player}} \\
         \hline
         \textbf{Explore} & \multicolumn{1}{c}{Low} & Unlimited recommendations (from 8 US YouTube categories, e.g., music, comedy) & Viral results - append “viral” to the user’s search query & \begin{itemize}
             \item Next video autoplays \item Related videos appear underneath 
         \end{itemize} \\
         \textbf{Focus} & \multicolumn{1}{c}{Medium} & Recommendations off by default (user can individually choose categories to turn on) & Relevant results - standard YouTube results for the user’s query & \begin{itemize}
             \item No autoplay
             \item No related videos
         \end{itemize}\\
         \textbf{Switch} & \multicolumn{1}{c}{High} & \multicolumn{3}{c}{Toggle lets the user switch between Focus Mode and Explore Mode. }\\
         \hline
    \end{tabular}    
    \label{tab5:threeversionsofswitchapp}
\end{table*}

Overall, about half of participants were actively trying to change or planning to do so shortly. A majority of these participants wanted more granular control over the type of content they consumed on YouTube, not just the ability to limit the quantity of time that current screen time tools provide. We aimed to address the needs of these participants through the design of the SwitchTube app.\\

Finally, the survey also served to screen and recruit participants into the next phase of our study, the field experiment. In therapy, the stages of change are often used to match the therapeutic process to the motivational stage of the client \cite{Norcross2011-mj}. Similarly, by enrolling only survey participants seeking to make a change in our field experiment, we could ensure that our technological intervention (SwitchTube) was likely to serve the needs of our study participants. 

\section{SwitchTube Field Experiment: Evaluating an Adaptable Commitment Interface}

Drawing upon prior work \cite{27_Lukoff2021-gm} led by many of the same authors of this current work and our survey results, we set out to understand how to design a mobile app that provides users with different levels of sense of agency over their YouTube experience. Two of the study authors, together with four students in an advanced-degree technology design program, used a design process of ideating, prototyping, building, and piloting before we evaluated SwitchTube in the field.

\subsection{Preparatory Design Work}

\noindent \textbf{Ideating.} We started by ideating 10 different design dimensions to manipulate user sense of agency over time spent, drawing upon attention capture dark patterns that have been previously proposed \cite{48_Zagal2013-iz,27_Lukoff2021-gm,Monge_Roffarello2022-qj, Monge_Roffarello2023-uu}. For example, Zagal et al. \cite{48_Zagal2013-iz} introduce ``playing by appointment'' wherein users are required to return to game within a fixed amount of time or else lose a reward. This led us to ideate the Time Pressure dimension, which ranged on a spectrum from no control to full control. Another dimension, Content Selection, varied from maximum to minimum temptation level. We then translated each of these dimensions into 23 sets of three concrete feature ideas each that ranged along this spectrum in terms of how much support they offered for user sense of agency (some dimensions inspired multiple feature sets). For example, for the Time Pressure dimension, we imagined video recommendations that expired if not watched within 30 minutes (low sense of agency), ones that expired within a day (medium sense of agency), and ones that were always available (high sense of agency). For Content Selection, we imagined a search algorithm that was tweaked to show results with a maximum entertainment level regardless of the user's actual query (low sense of agency), one version that showed both entertaining and relevant results (medium sense of agency), and one version with only relevant results (high sense of agency). As a group, we then scored these feature sets in terms of expected impact, novelty, and technical feasibility. \\

\noindent \textbf{Prototyping.} Paper mockups for the seven highest-scoring feature sets were evaluated in 13 co-design sessions with YouTube users, as described in our prior work \cite{27_Lukoff2021-gm}. For example, Figure \ref{fig4:fig_prototypes} shows the prototype for Content Selection with three different versions of search results. We initially anticipated building three different versions of SwitchTube along a spectrum in terms of their support for sense of agency (low, medium, high) to find a “Goldilocks” level of control as has been suggested in prior work on lockout mechanisms \cite{22_Kim2019-pq,29_Lyngs2022-ug}. However, co-design sessions with YouTube users revealed that rather than having a stable preference at all times, users wanted different levels of control for different situations. For example, when they had a specific intention in mind, they preferred a search-first interface, whereas when they just want to relax or pass the time, they preferred a recommendations-first interface \cite{27_Lukoff2021-gm}. Taking these findings into account, we instead designed two versions that support different levels of sense of agency, and a third version where users could switch between the two (hence the name SwitchTube). It also prompted us to consider that Switch (rather than Focus) might offer the highest sense of agency of all of the versions, a hypothesis that we test in this work. \\

We created an interactive mockup of our complete SwitchTube design in Figma and conducted usability testing with 4 participants, all university students who were active YouTube users. Participants completed four tasks, which helped us identify a number of smaller usability issues. One of the usability testing participants said they would like to use the low-agency version to \textit{“explore viral content”} and the high-agency version to \textit{“focus on my goal,”} which led us to call the two versions “Explore” and “Focus.” While labeling the two versions in this way could lead study participants to form preconceived notions of how to use that version (as opposed to say, “Version A” and “Version B”), we decided that this was worthwhile because it would make it easy for participants to recall the two versions in the exit survey and interview and seemed unlikely to signal to participants that they \textit{should} necessarily prefer one version over the other.\\

Table \ref{tab5:threeversionsofswitchapp} shows an overview of the three different versions of the final SwitchTube study app and their features. A screenshot of the homepage of each of the three versions is shown in Figure \ref{3versions_switchtube}. A short video introduction and captioned screenshots of the entire app are also available on the Open Science Framework: \href{https://osf.io/z735n/}{\color{blue}{https://osf.io/z735n}}. We refer to the three versions of the app as the Explore Version, Focus Version, and Switch Version, and the two toggle options \textit{within}\:the Switch Version as Explore Mode and Focus Mode.
\begin{figure*}
    \centering
    \includegraphics[width=0.9\textwidth]{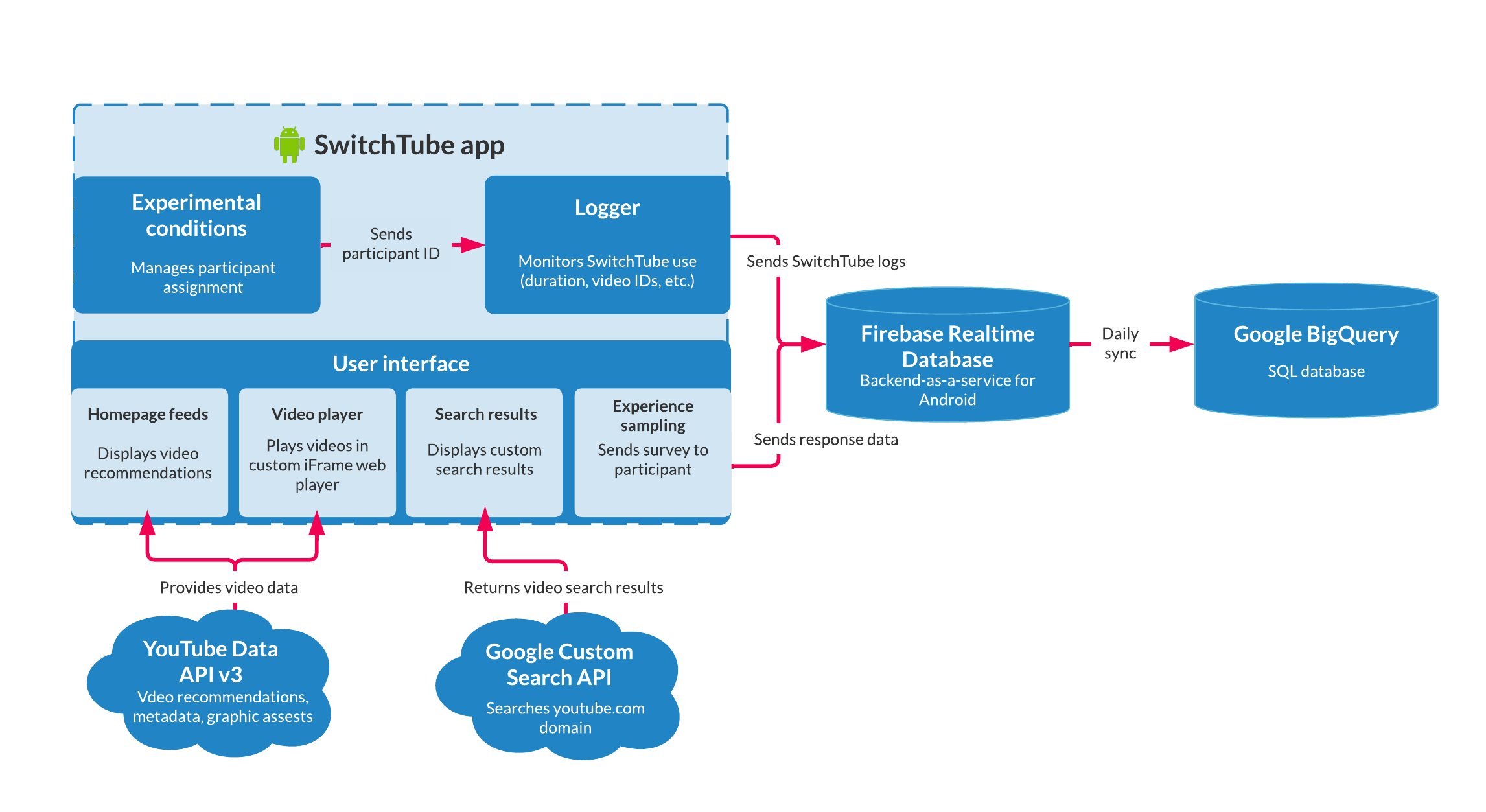}
    \caption{The technical implementation of the SwitchTube study app for Android.}
    \label{fig5:technicalimplementation}
    \Description{An image showing the structure of the SwitchTube app. YouTube Data API v3 is connected to Homepage feeds and Video player, and Google Custom Search API is connected to search results. Experimental conditions send participant ID to Logger. Both logger and Experience sampling sends data to the Firebase Realtime Database. The Firebase Database daily syncs with Google BigQuery.}
\end{figure*}

Our aim for Focus was to support the user's specific intention for visiting the app if they had one (e.g., learning how to cook a turkey), whereas Explore was designed to maximize distractions that would take them away from their original intention. In doing so, we expected that sense of agency (the user's experience of being the initiator of their actions) would be supported in Focus and diminished in Explore. To this end, we appended ``viral'' to every search query submitted in the Explore Version. Our goal was not to fill Explore with viral content per se, but rather to add noise \textit{and} temptations to the user's search results. To simply add noise, we could have appended \textit{any} term to the user's search query (e.g., ``zebras''), but our internal testing of several different terms (e.g.,  ``entertaining,'' ``funny,'' and  ``creative'') suggested that ``viral'' was the most effective at returning results that were also tempting.\\

Homepage video recommendations in SwitchTube were not personalized due to restrictions of the YouTube Data API (personal watch history and recommendations are difficult-to-access due to understandable privacy concerns). We further address the absence of personalized recommendations in the discussion section. Instead, homepage recommendations were drawn from the most popular videos in different YouTube categories (e.g., music, comedy) for the U.S. region. These are the same non-personalized recommendations that are displayed in YouTube's own categories. In the Explore Version, the homepage featured an unlimited scroll of these videos and the video player showed related video recommendations below the video that was currently playing and autoplayed the next related video. In the Focus Version, the homepage hid recommendations by default and related videos and autoplay were removed by design. \\

\noindent \textbf{Building.} An illustrated software architecture model for the SwitchTube study app on Android is shown in Figure \ref{fig5:technicalimplementation}. The app assigned participants to experimental conditions and used a logger to monitor information about how participants used the app. The user interface had homepage video feeds, a video player, and search results. These were populated with data pulled from the YouTube Data API and the Google Custom Search API. Finally, the app conducted experience sampling, which we built as a custom system. All of this data was sent to the Firebase Realtime Database and then synced with Google BigQuery to allow for custom views and further analysis.

One particular challenge we encountered was a severely limited quota for the YouTube Data API, which we needed to populate the video recommendations on the homepage and video player (related videos and autoplay) and return search results. When we built SwitchTube, YouTube restricted developers to a default quota of 10,000 per day, whereas default quota at the time of previous research had been 1 million at the time of previous YouTube research \cite{19_Hiniker2017-bj}. As a result, we quickly maxed out our quota in our testing of the app (e.g., a single search has a quota cost of 100). We tried the official form for requesting an increased quota, but received no response. Drawing upon our privileged position, we contacted multiple personal connections at Google in managerial positions who were also unable to get the YouTube Data API team to grant our request. In the end, we were forced to integrate a second API into SwitchTube (the Google Custom Search API), which we could pay for and use to populate search results, but it cost us considerable time and effort to do so. Two years later and after we had finished our deployment study, we received an email that YouTube has finally launched an official YouTube Researcher Program with expanded access to their Data API\footnote{https://research.youtube/}. We hope our report of this barrier lends support to the regulatory push to require large technology companies to provide researchers with greater access to audit and redesign their algorithmic systems for digital wellbeing. \\

\noindent \textbf{Piloting.} Our research team internally piloted the SwitchTube study app on a variety of Android devices over eight weeks. This again identified countless usability issues, from the font size of the experience sampling prompts to missing log data, that we resolved in the next version. We then recruited four students, all active YouTube users, from outside the research team for external piloting, which identified still further issues about study procedures, but also confirmed that the app was ready for deployment. We note that these participants identified several usability issues that we simply decided not to fix (e.g., when the phone was rotated horizontally the video had to reload). Our goal was not to rebuild a user experience as seamless as YouTube itself (which would have required a Herculean effort), but rather to develop a proof-of-concept system that would be acceptable enough for participants that they would engage with it sufficiently to address our research questions \cite{Fogarty2017-gi}. 

\subsection{Pre-Registered Hypotheses}
Our third research question asks how adaptable commitment interfaces influence user experience. In line with this question, we posed several specific hypotheses. Following the best practices of the open science movement, we pre-registered these \textit{before} examining the data: \href{https://osf.io/sevfd}{\color{blue}{https://osf.io/sevfd}}. This helped us think through our study protocols in advance and guard against the the natural temptation of hypothesizing after the results are known (HARKing) \cite{9_Cockburn2018-ag}. As noted in the pre-registration, in addition to this \textit{confirmatory} \textit{analysis} with pre-registered hypotheses we also planned to conduct \textit{exploratory} \textit{analyses} of the log data from the app, such as time spent in the different versions, but to use only descriptive statistics for this purpose. \\

In general, our pre-registered hypotheses tested whether or not the Switch Version (an adaptable commitment interface) provides the `best of both worlds' across measures for sense of agency, satisfaction, and personal goal alignment. All of our hypotheses were tested based on measuring the mean per participant rating (1-7) of experience sampling mechanism (ESM) responses for these metrics. \\

\noindent \textbf{H1: User Sense of Agency.} Our first set of hypotheses (H1a-H1c) addressed user \textbf{sense of agency}, which prior work suggests is at the center of user concerns with social media \cite{Baumer2018-gv}. Our expectation was: Switch > Focus > Explore, which corresponds to 3 pairwise comparisons:
\begin{itemize}
    \item H1a: The mean rating will be higher for Focus than Explore.
    \item H1b: The mean rating will be higher for Switch than Explore.
    \item H1c: The mean rating will be higher for Switch than Focus.
\end{itemize}
The features in Focus and Explore were based on our prior research into how the features of YouTube affect user sense of agency \cite{27_Lukoff2021-gm}. As Switch lets users toggle between the Focus and Explore interface, we expected that this additional option would further increase user sense of agency.\\

\noindent \textbf{H2: Satisfaction.} Our second set of hypotheses (H2a-H2c) addressed user satisfaction, as in the short-term pleasure that users derive from social media apps. Our expectation was: Switch > Explore > Focus, again corresponding to 3 pairwise hypotheses that follow the same pattern as H1. In our previous study of YouTube \cite{27_Lukoff2021-gm}, users reported that homepage recommendations often provided short-term satisfaction, but Focus hides these by default. We expected Switch might provide a useful option to avoid recommendations at times when they are not wanted. \\

\noindent \textbf{H3: Goal Alignment.} Our third set of hypotheses (H3a-H3c) addressed personal goal alignment, as in how well app use aligned with the user’s long-term goals for use. Our expectation was: Switch > Focus > Explore, which again implies 3 pairwise comparisons. This is because our previous work found that that search often supports YouTube users' personal goals \cite{27_Lukoff2021-gm}, but Explore minimizes the search option and adds distracting temptations to the results. On occasions where recommendations might actually better support the user than search (e.g., as in when survey participants said their goal was to find new or diverse content to watch), Switch would also provide that option.

\subsection{Methods}
\textbf{Recruitment.} We screened the 606 participants from our survey for the following three inclusion criteria:
\begin{enumerate}
    \item Action or preparation stage of change with regards to their YouTube use (48\% of survey participants met this criterion).
    \item Own an Android smartphone with operating system version 6.x - Marshmallow or higher. This was because the study app did not support older versions (87\% of survey participants met this criterion).
    \item Spend a minimum of 10 or more minutes per day on the YouTube mobile app, according to self-estimate (75\% of survey participants met this criterion). This was to ensure that participants already had a regular habit of watching videos on mobile, making it more natural for them to use SwitchTube.
\end{enumerate}

\noindent This left us with 146 survey participants who were eligible to also become experiment participants. Given that a prospective power analysis for ESM studies requires an estimate of effect size that is difficult to obtain for a novel technology, we instead followed Berkel et al.’s guidance and informed our target number of participants using local standards in the HCI community \cite{3_Berkel2017-gn}, where the median is 18 participants and the mean is 53 \cite{6_Caine2016-am}. Since we wanted to be able to detect differences between conditions with a high degree of confidence using frequentist hypothesis testing, we set a target of having 45 participants complete the field experiment. \\

\setcounter{table}{6}
\begin{table*}
    \centering
    \caption{SwitchTube Study Timeline}
    \begin{tabular}{|l|L|L|L|L|L|L|}
    \hline
    \rowcolor{lightblue}
    & \textbf{Start} & \textbf{Week 1} & \textbf{Week 2} & \textbf{Week 3} & \textbf{Finish} & {\textbf{Subset of participants}} \\
    \hline
    \cellcolor{lightblue}
    Study App & Install the app & Explore or Focus & Explore or Focus &	Switch & Uninstall the app &  \\
    \hline
    \cellcolor{lightblue}
    Activities & Entrance Survey & \multicolumn{3}{K|}{Each week, use the app on 3+ days for a total of at least 30 minutes. Answer the experience sampling questions.} & Exit Survey & Exit interview \\
    \hline
    \cellcolor{lightblue}
    Incentive & \$5 & \$15 & \$30 & \multicolumn{2}{M|}{\$50\newline (for Week 3 and Exit Survey)} & \$20 \\
    \hline
    \end{tabular}        
    \label{tab7:SwitchTubeStudyTimeline}
\end{table*}

\begin{table*}
    \centering
    \caption{Experience Sampling Questions}
\begin{tabular}{lp{10.5cm}p{4cm}}
         \hline
         \rowcolor{Gray}
         & \multicolumn{1}{c}{\textbf{Question}} & \multicolumn{1}{c}{\textbf{Scale}}\\
         \hline
         \textbf{Sense of agency} & For this SwitchTube use, how much did you feel out of or in control? & 1=very out of control\newline 7=very in control \\
         \textbf{Satisfaction} & For this SwitchTube use, how much did you feel dissatisfied or satisfied? & 1=very dissatisfied\newline 7=very satisfied \\
         \textbf{Goal alignment} & For this SwitchTube use, how much did it conflict with or support your personal goals? & 1=very in conflict \newline 7=very supported \\
         \hline
    \end{tabular}
\label{tab8:experiencesamplingquestions}
\end{table*}
We invited eligible survey participants to participate in small batches until we approached our target. In the invitation to the study and again upon installing the study  app, participants were informed that the research team would monitor and analyze their activity in the study app, including their searches and the titles of the videos they watched.\\

\noindent \textbf{Demographics.} A total of 46 participants completed the experiment (see demographics in Table \ref{tab6:46Demographics}). We happened to oversample Asian, Black, and young people relative to the general U.S. population \cite{43_United_States_Census_Bureau_undated-re}.
\setcounter{table}{5}
\begin{table}[H]
    \centering
    \caption{Demographics of the 46 field experiment participants}
    \begin{tabular}{|>{\columncolor[gray]{0.85}} l|p{5.5cm}|}
    \hline
         Gender identity & Man (54\%), Woman (46\%), Non-binary (0\%), Prefer not to say (0\%)  \\
         Age range & 18-24 (35\%), 25-34 (39\%), 35-44 (17\%), 45-54 (4\%), 55+(4\%) \\
         Education & High school (35\%), Associate degree (13\%), Bachelor's degree (37\%), Advanced degree (15\%) \\
         Race/ethnicity & White (43\%), Asian (26\%), Black (20\%), Hispanic (7\%), 2 or more races (4\%) \\
         \hline
    \end{tabular}
    \label{tab6:46Demographics}
\end{table}

\noindent \textbf{YouTube Use.} Field experiment participants spent a median of 140 minutes per day (interquartile range: 120-240) on YouTube across all devices in the week prior to the survey (self-estimated). Of this time, participants estimated they spent a median of 63\% (interquartile range: 40-84\%) in the mobile app. We again multiplied time on all devices with the percentage spent on mobile, for each participant, to find that field experiment participants spent a median of 87 minutes per day in the YouTube mobile app. This is considerably higher than the median of 34 minutes per day spent in the app by all survey participants, indicating that those who were invited and participated in the field experiment were heavier YouTube users.\\

\noindent \textbf{Procedures}. As shown in Table \ref{tab7:SwitchTubeStudyTimeline}, participants completed an entrance survey, one week of use of each of the three versions of the SwitchTube app (Explore, Focus, Switch), an exit survey, and, for a subset of participants, an exit interview. In the entrance survey, participants completed additional questions about the nature of their YouTube use and received instructions for installing the SwitchTube study app on their Android phone from the Google Play store.\\

Upon installing SwitchTube, participants were assigned to start in either Explore or Focus following a counterbalanced assignment. Although it risked introducing ordering effects, we decided against also counterbalancing the Switch condition. Instead, Switch always came last so that we could understand when and why participants choose to toggle between Explore and Focus (RQ2) after having experienced each for a week. In each week, participants were required to use the app for 3 or more days for a total of at least 30 minutes. If participants did not meet these requirements, they were disqualified from further participation, but still compensated for their participation to that point. \\

The SwitchTube app collected both objective and subjective data. Objective data included logs of time spent, searches made, videos watched, and the source of watched videos (e.g., homepage recommendations). In terms of subjective data, participants were experience sampled using the three questions in Table \ref{tab8:experiencesamplingquestions}. Conceptually, we wanted to capture an understanding of how the different versions influenced sense of agency, as well as satisfaction in the sense of short-term pleasure and goal alignment in the sense of long-term personal goals. Unfortunately we could not find validated scales that were short enough to be suitable for ESM, but we tested our wording for clarity in our piloting. This led us to clarify that we wanted participants to answer about this particular session of use \textit{(“For this SwitchTube use”)} rather than for their use of SwitchTube as a whole.\\

In terms of timing, a prompt appeared with these three questions on the participant’s phone when the following conditions were met:
\begin{enumerate}
    \item The participant had not already responded in the past hour;
    \item The participant had used the app for at least 30 seconds;
    \item The app went into the background (e.g., the user exited the app to the phone’s home screen or they switched to another app).
\end{enumerate}
If the participant did not respond within one minute, the prompt disappeared.\\

After completing one week each in Explore, Focus, and Switch, participants completed an exit survey. In the exit survey, participants were shown screenshots of the homepage, search results page, and video player in Explore and Focus as a reminder and answered which they preferred and why. Participants then explained when and why they switched between versions of the app. Finally, they answered which version of the app they preferred and why. \\

Exit interviews were conducted remotely over Zoom with a subset of participants using a method called data-driven retrospective interviewing \cite{39_Seif_El-Nasr2015-ou}. Using screen share, participants were shown counts, tables, and visualizations from their own log data, e.g., time spent in the app, occasions when they switched between versions, and their ESM ratings, and asked questions intended to elicit the “why” behind their behaviors. For example, we asked:
\begin{quote}
    In the Switch version, you switched between the Focus and the Explore interface 17 times. Can you look at the table below, choose a couple of examples, and describe why you switched at that time?
\end{quote}
\noindent We also retrieved the original change that participants wanted to make to their YouTube use from the survey and asked them whether the different versions of SwitchTube supported that goal. A total of 16 participants were interviewed, at which point we believe we reached data saturation with regards to our research questions. Interviews lasted about 45 minutes each.\\

Participant incentives were backloaded to encourage participants to complete the entire study, allowing us to compare their experience between conditions. This meant: \$5 for the entrance survey,  \$15 for week 1 of app use, \$30 for week 2, \$50 for week 3 and the exit survey, and \$20 for the exit interview. To protect data privacy, we assigned each participant a unique identifier (e.g., 446565) that was associated with their usage data. We connected this data to the participant’s personally identifiable information (e.g., contact information) only for the exit interviews, where we presented participants with a personalized summary of their usage. This research was approved by the University of Washington IRB.

\subsection{Data Analysis}
Log data were analyzed in an exploratory fashion and are presented as descriptive statistics. Experience sampling data were analyzed according to our pre-registered hypotheses. Exit survey and exit interview data were analyzed together by three of the authors, who conducted a codebook thematic analysis that addressed our research questions \cite{4_Braun2019-af}. We first read through all of the data and added initial codes using Delve, a tool for collaborative qualitative coding. Initial codes were refined and consolidated through group discussion, leading to a final codebook which was applied to the data. For example, one code was “within-session switching,” which was for cases in the Switch Version where participants described a use case of toggling between Explore Mode and Focus Mode during a single visit, for example:
\begin{quote}
    I actually would switch between the two regularly, but specifically to use focus as a search bar and explore as a way to find new content I was not thinking of. (P41)
\end{quote}
\noindent Each code and its supporting quotes was then moved to Miro, a tool that that we used to conduct affinity diagramming, in which codes were clustered according to our research questions. Through this process, we found that there were four distinct use cases for switching. We then drew upon these codes and their associated quotes to write four analytical memos: switching behavior (for RQ2) and sense of agency, satisfaction, and goal alignment (for RQ3), which form the basis of our results.

\section{Results and Analysis}
Our log data, experience sampling, and interview data allowed us to triangulate the experience of 46 participants in our study using both objective and subjective measures. 

\subsection{App Usage Data}
\begin{figure*}
    \centering
    \includegraphics[scale=0.15]{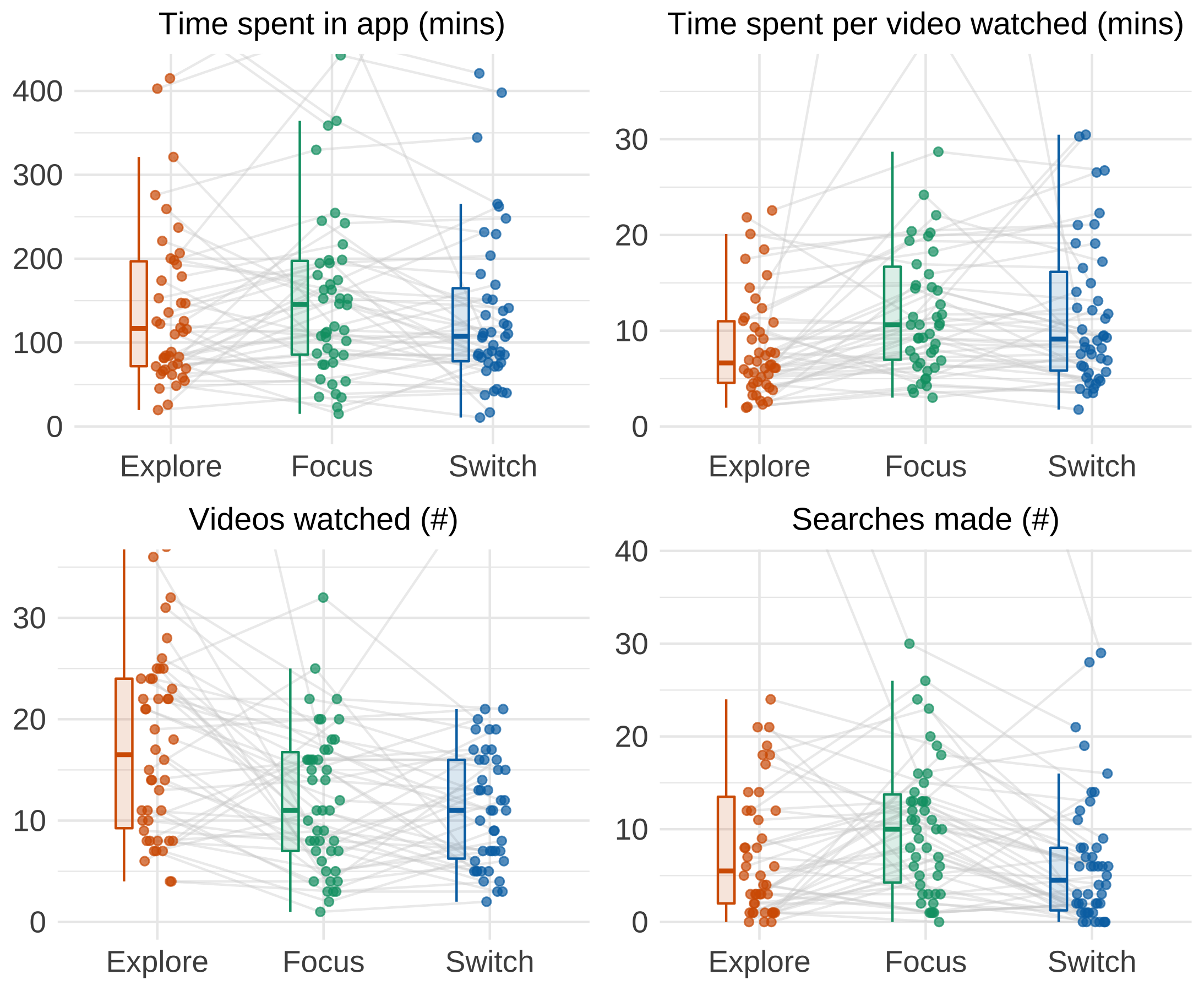}
    \caption{Log data from our deployment study with 46 participants for the three versions of the SwitchTube app: Explore, Focus, and Switch. Participants spent more time in Focus than Explore. Participants also spent more time on each video that they watched (or started) in Focus, although they actually watched more videos in Explore. Participants in Focus also made more searches than in the other versions. For a detailed explanation of the format used for this figure, which is repeated in several subsequent figures, see the footnote. \protect\footnotemark}
    \label{fig6:logdata}
    \Description{A figure with a total of four visualizations. The first visualization is about the time spent in the app (in minutes). In that visualization, Explore is slightly lower than Focus, and Focus is higher than both Switch and Explore. The second visualization is about the time spent per video watched (in minutes). In that visualization, Explore is lower than both Focus and Switch, and Switch is lower than Focus. The third visualization is about the videos watched. In that visualization, Explore is higher than Focus and Switch, and Focus and Switch are about the same, but Focus has a wider range. The last visualization is about the number of searches made, where explore  is lower than Focus, and Switch is lower than Explore.}
\end{figure*}

In terms of usage of SwitchTube, there were notable differences and non-differences in how participants engaged with the three versions of the app for one week each (Figure \ref{fig6:logdata}). Our analysis of usage focuses primarily on the comparison between Explore and Focus, before turning to when and why participants used the toggle in Switch (RQ2) and how each version influenced user experience (RQ3).

\textbf{Time spent.} The median time spent using each version of the app for a week was as follows:

\begin{table}[H]
    \centering
    \begin{tabular}{LLL}
        \rowcolor{Gray}
        \multicolumn{3}{c}{\textbf{Time spent in the app per week (minutes)}} \\
        Explore & Focus & Switch \\
        117 & 145 & 107
    \end{tabular}
\end{table}

\noindent On the one hand, participants spent considerable time with our study app, far beyond the 30 minutes per week that we asked of participants. At most, one participant even spent 752 minutes (12.5 hours) in a week in the app in Switch. On the other hand, given the high usage of YouTube that our participants reported (a median of 87 minutes per day on mobile), SwitchTube still likely represented only a fraction of their overall time spent watching YouTube videos. Surprisingly, the time spent in Focus actually exceeded the time spent in Explore and Switch, whereas we had anticipated that the opposite would be the case because of the lack of recommendations in Focus.\\

\textbf{Time spent per video watched.} The median time spent per video that participants started to watch (regardless of whether or not they finished it) was as follows:
\begin{table}[H]
    \centering
    \begin{tabular}{LLL}
        \rowcolor{Gray}
        \multicolumn{3}{c}{\textbf{Time spent per video watched (minutes)}} \\
        Explore & Focus & Switch \\
        6.6 & 10.6 & 9.1
    \end{tabular}
\end{table}

\noindent Participants spent more time on each video that they started in Focus, which partially explains why the overall time spent was greater in that version of the app. \\

\textbf{Videos watched.} The median number of videos that participants watched over the course of a week with each version was as follows:
\begin{table}[H]
    \centering
    \begin{tabular}{LLL}
        \rowcolor{Gray}
        \multicolumn{3}{c}{\textbf{Videos watched (\#)}} \\
        Explore & Focus & Switch \\
        16.5 & 11 & 11
    \end{tabular}
\end{table}
\noindent Participants started to watch more videos in Explore than in the other versions, even though they spent less overall time in Explore. In other words, participants in Explore started more videos that they didn’t finish as their selections were driven by recommendations rather than content than they had specifically searched for.  The videos that participants selected in Explore were also slightly shorter (median 8.8 minutes) than in Focus (10 minutes) and Switch (11.9 minutes). \\

\textbf{Searches made.} The median number of searches that participants made was as follows:
\begin{table}[H]
    \centering
    \begin{tabular}{LLL}
        \rowcolor{Gray}
        \multicolumn{3}{c}{\textbf{Searches made (\#)}} \\
        Explore & Focus & Switch \\
        5.5 & 10 & 4.5
    \end{tabular}
\end{table}
\noindent Participants made more searches in Focus, suggesting that (a) the lack of recommendations led to more searching and/or (b) participants preferred the relevant results in Focus over the viral results in Explore.\\

\textbf{Time spent browsing.} The median percentage of time spent browsing in the app (any time spent in the app when no video was playing), was as follows:
\begin{table}[H]
    \centering
    \begin{tabular}{LLL}
        \rowcolor{Gray}
        \multicolumn{3}{c}{\textbf{Time spent browsing (\%)}} \\
        Explore & Focus & Switch \\
        49\% & 46\% & 38\%
    \end{tabular}
\end{table}

\noindent Participants in Explore spent 3 percentage points more of their time in SwitchTube browsing than they did in Focus and 11 percentage points more than they did in Switch. This is likely because there were more recommendations to browse in Explore. \\

\begin{figure*}
    \centering
    \includegraphics[scale=0.18]{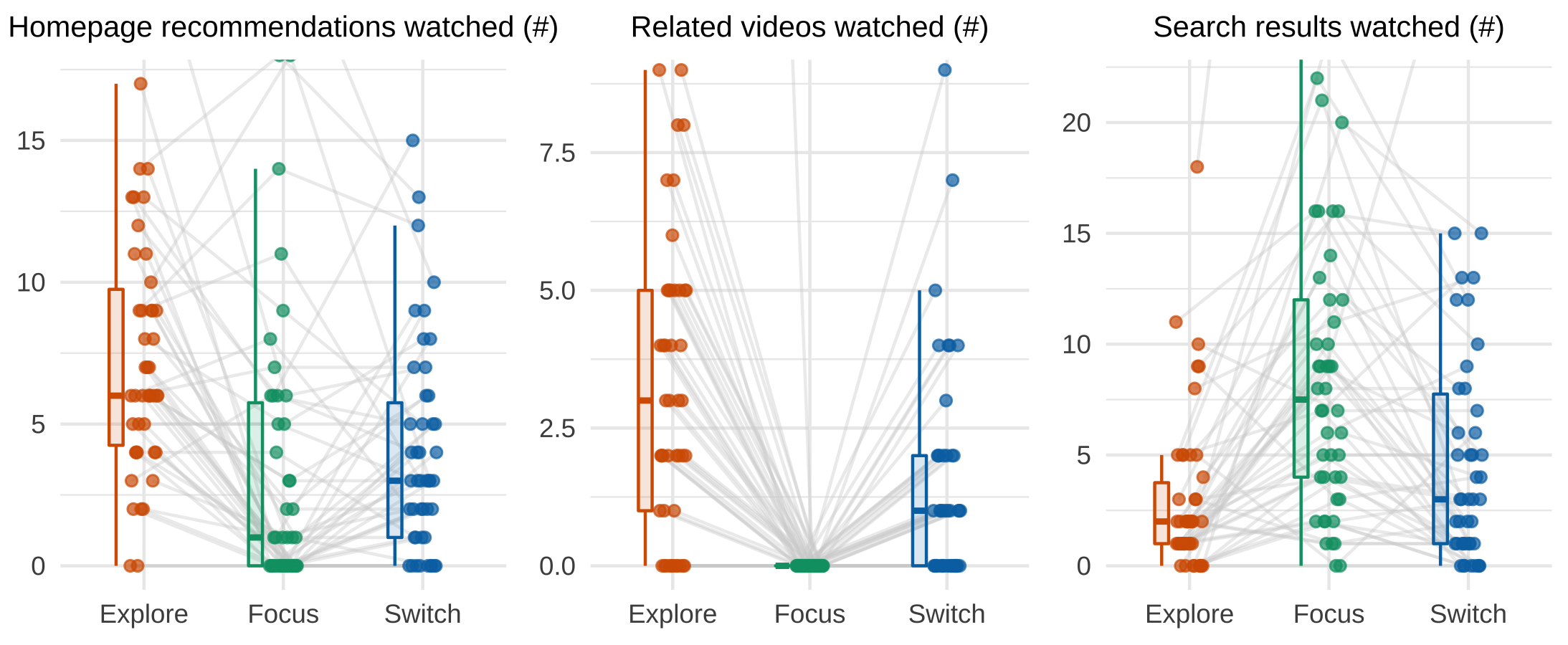}
    \caption{Videos watched from three different sources: homepage recommendations, related videos (which appeared underneath the video that was currently playing), and from searches. Video watching was driven by recommendations in Explore and by searches in Focus. In Switch, there was a balance between the two.}
    \label{fig7:video3sources}
    \Description{Three visualizations. The first visualization is about the number of home page recommendations watched, where Explore is the highest, and Focus is lower than both Switch and Explore. The second visualization is about the number of related videos watched, where Explore is higher than Switch, and Focus is at 0. The third visualization is about the number of search results watched, where Explore is lower than both Focus and Switch, and Focus is higher than Switch.}
\end{figure*}

\textbf{Videos watched by source.} Participants could watch videos from the homepage recommendations, from related videos (which appeared underneath the video that was currently playing), or from searches (Figure \ref{fig7:video3sources}). From homepage recommendations, participants watched far more in Explore (median: 6 videos) than in Focus (median: 1), indicating the strong influence of an interface where the user has to opt-in to recommendations. By contrast, Explore and the current interface of the YouTube mobile app show recommendations without even providing an opt-out choice. From related videos, participants watched a median of 3 videos in Explore, whereas these were completely unavailable in Focus. From search, participants watched far fewer videos in Explore (median: 2) than in Focus (median: 7.5). Across these three video sources, Switch was in the middle between Explore and Focus suggesting that participants used features from both versions of the app, rather than just keeping the toggle on one mode.\\

\textbf{Videos watched by kind.} 
In Focus, participants tended to consume videos that all stemmed from the same search and were thus closely related: for instance, P21 played ``15-minute morning power yoga'' \cite{SarahBethYoga2019-qr} followed by ``15-minute hands free morning yoga'' \cite{SarahBethYoga2019-io}. By contrast, in Explore, the same participant jumped from an ASMR-relaxation video to a celebrity cooking challenge to a popular hip-hop music video, consuming a great diversity of content. In Explore, viewing was driven in large part by homepage recommendations of the most popular videos in the U.S., so videos tended to appeal to a broad audience (e.g., ``If You Can Carry \$1,000,000 You Keep It!'' \cite{MrBeast2021-wd}) and often included creators who are celebrities on YouTube (e.g., MrBeast) and beyond (e.g., Ed Sheeran). By comparison, participants in Focus often viewed more niche videos, such as one on opening moves in chess that is unlikely to be recommended on the homepage as one of the most popular videos in the U.S. \\

We make available an open dataset of 2063 video play events, including the title of the video, the version of the app (Focus, Explore, or Switch), and the source of the video (search, related videos, or homepage recommendations) via the Open Science Framework: \href{https://osf.io/z735n/}{\color{blue}{https://osf.io/z735n}}. To protect the privacy of study participants, no participant ID is included. While outside of the scope of this paper, future work could analyze this data to identify further differences in the nature of the content consumed in the different versions of SwitchTube that might inform the design of content algorithms for digital wellbeing. \\


\subsection{The SwitchTube User Experience}
We triangulated data from multiple sources to understand how participants experienced SwitchTube. Log data served as an objective measure of feature use, while experience sampling, the exit survey and interviews revealed contextual information and the subjective experience behind the usage.\\

\subsubsection{Why and When People Switched (RQ2)}

\footnotetext{The boxplot shows the median, the interquartile range (Q1-Q3) and, with the top whisker, Q3 + 1.5x the interquartile range. Each dot represents one of the 46 study participants and the gray lines show how that participant’s data changed between versions of the app. To zoom in on the vast majority of the data, we calculated the 95th percentile for all three versions, took the maximum value, and cropped the top of each figure at that point; the gray lines going off the top indicate outliers that extend above the top of the figure. The Explore and Focus conditions were counterbalanced in the actual experiment, but are always displayed Explore first, Focus second in this figure. Switch always came last, as discussed in the paper.}

In Switch, users could toggle between Explore and Focus Mode. Almost all users made use of this feature, making a median of 8.5 switches during their week in the app (interquartile range 4-13). Our analysis found that \textbf{four distinct use cases} captured switching behavior and motivation:

\paragraph{\textbf{Use Case 1 — Curiosity about feature differences}}
When they first opened Switch, participants switched between the Explore and Focus Modes in rapid succession. Although they had already experienced each of these individually for a week, they were still curious about the differences between the two: \textit{“I went back and forth a little bit just to see if anything was added to either one of them, or which one offered more” (P7)}. This curiosity-driven switching not only satisfied an itch to understand the differences, but also facilitated preference discovery, i.e., helping participants decide which version was a better fit for them.
\begin{figure*}
    \centering
    \includegraphics[width=\textwidth]{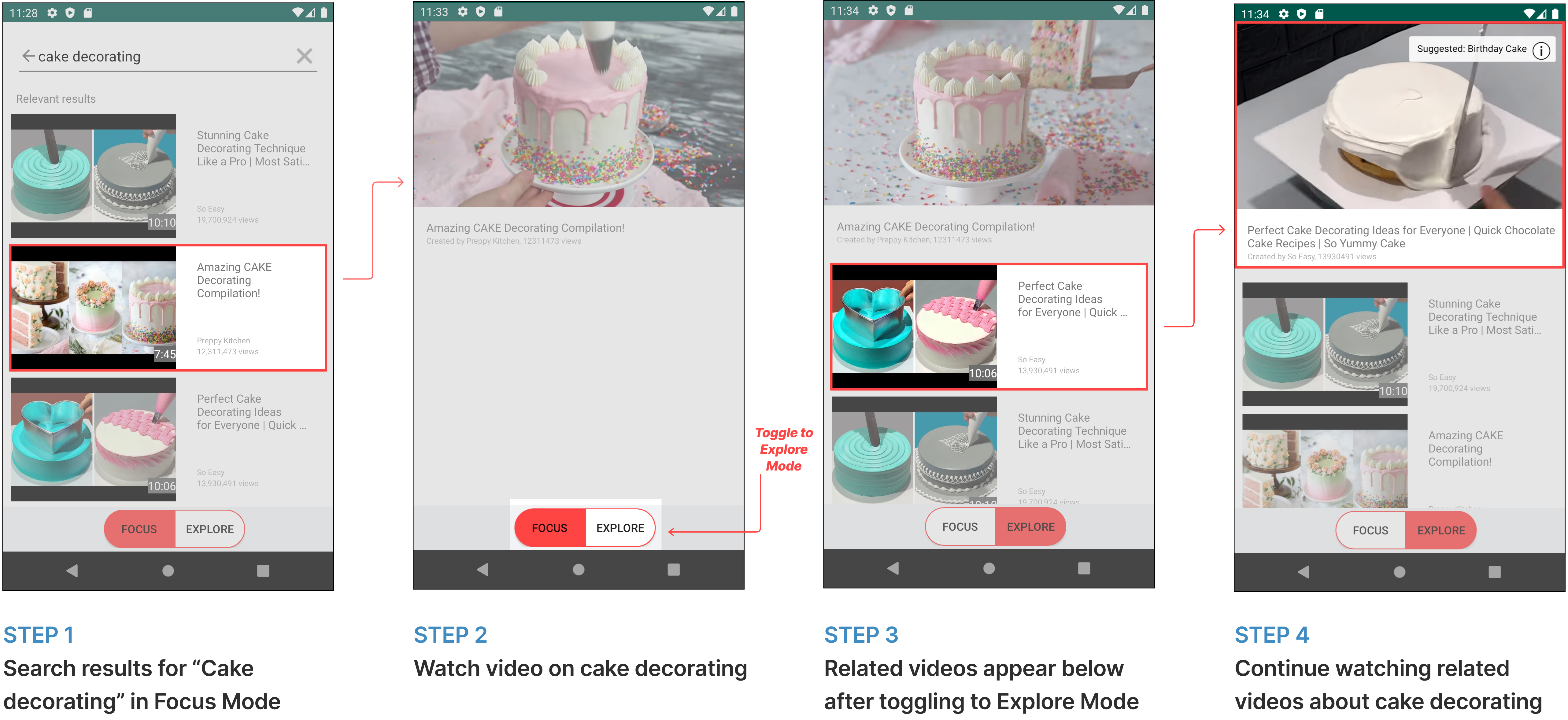}
    \caption{In this example of Use Case 4, recreated based on one of our exit interviews, P3 started with a search in Focus and then toggled to Explore to see related videos. By switching within a single session of use like this, participants were able to perform a search in line with their initial intention without getting distracted, but then still explore related content from there.}
    \label{fig8:Case4}
    \Description{An image showing the steps of watching related videos with red arrows indicating what users clicked on. The user first clicked on a video in Focus mode. When they are watching the video, there were initially no recommended videos showing underneath. Once the user toggled to the Explore mode, the interface will start showing recommended videos for users to watch.}
\end{figure*}

\paragraph{\textbf{Use Case 2 — Curiosity about new content}}
Participants sometimes switched from Focus to Explore Mode to see recommendations:
\begin{quote}
   I would occasionally switch to Explore to see if there was anything popular—it was how I discovered The Weeknd released a new song! But for the most part I preferred to stay in Focus. (P1)
\end{quote}
\noindent This use case aligns with one of the top goals that survey participants had: watching more new or diverse content. Sometimes recommendations inspired participants to search in a new direction: \textit{“While I do like searching for content myself, sometimes I do get bored and looking at genres reminds me of things I would also like to search for”} (P36). In these cases, switching from Focus to Explore Mode satisfied a need for novelty.\\

\paragraph{\textbf{Use Case 3 — Between-session switching: addressing situational needs}}
Another common use case was when a participant had a particular intention for their session of use and toggled to the most suitable interface: \textit{``If I'm searching for something like ‘flat tire’ I'd go with Focus because that seems to be more educational. Explore is definitely more entertainment''} (P38). Similarly, another participant said,
\begin{quote}
    With Explore showing the most watched content, I often found what I wanted easier with it. If I was however watching something for educational purposes I always used Focus first, finding myself less prone to wasting time watching senseless videos. (P12)
\end{quote}
\noindent In this use case, the adaptable commitment interface of Switch was effective at holding the user to their commitment.

For another participant, Focus also made them realize how strongly design patterns influenced their behavior:
\begin{quote}
    Focus was better for restricting my viewing and keeping with my restrictive schedule. In Focus, the app would let me select a video and after it was done it would end with no autoplay directly afterwards. What I learned was that autoplay was a big reason for continual watching; far more than I ever knew. (P10)
\end{quote}
\noindent Other participants appreciated how Focus limited \textit{“distractions,”} \textit{“unintended videos,”} and \textit{“losing track of time.”} This pattern of use accorded with our prior research: users come to YouTube with many different intentions, so recommendations are sometimes helpful. However, the current interface of the YouTube app aggressively promotes new content discovery, even when that is not the user’s intention \cite{27_Lukoff2021-gm}.\\

\paragraph{\textbf{Use Case 4 — Within-session switching: start focused, explore from there}}
A final and unanticipated use case was that participants frequently toggled between Focus and Explore within the same session, as shown in Figure \ref{fig8:Case4}.
\begin{quote}
    I used Focus to search for a particular video and when I was done watching I would turn on Explore sometimes, when I felt like looking for recommended results underneath. (P13)
\end{quote}
\noindent In this use case, participants started in Focus Mode to avoid distraction from homepage recommendations. Then they searched for and clicked on a relevant result. Once the video started or finished playing, they then toggled over to Explore Mode to reveal related videos. In this way, they could feed the system with their initial intention without getting distracted, and then explore from there. With an understanding of app usage data and switching behaviors, we now turn to how SwitchTube influenced user experience measures.

\subsubsection{\textbf{H1: Sense of Agency}}
In order to analyze the experience sampling ratings, starting with sense of agency, we first had to select an appropriate statistical method. Since our data were derived from an ordinal scale, determining whether or not the means of each participant ratings are normally distributed was important. We performed a Shapiro-Wilk test that showed that the distribution of participant means in the Switch condition departed significantly from normality (W = 0.96, p-value < 0.01). Based on this outcome, we use non-parametric statistics in all subsequent analyses of ESM ratings.\\ %

\begin{figure}[htbp]
    \centering
    \includegraphics[scale=0.10]{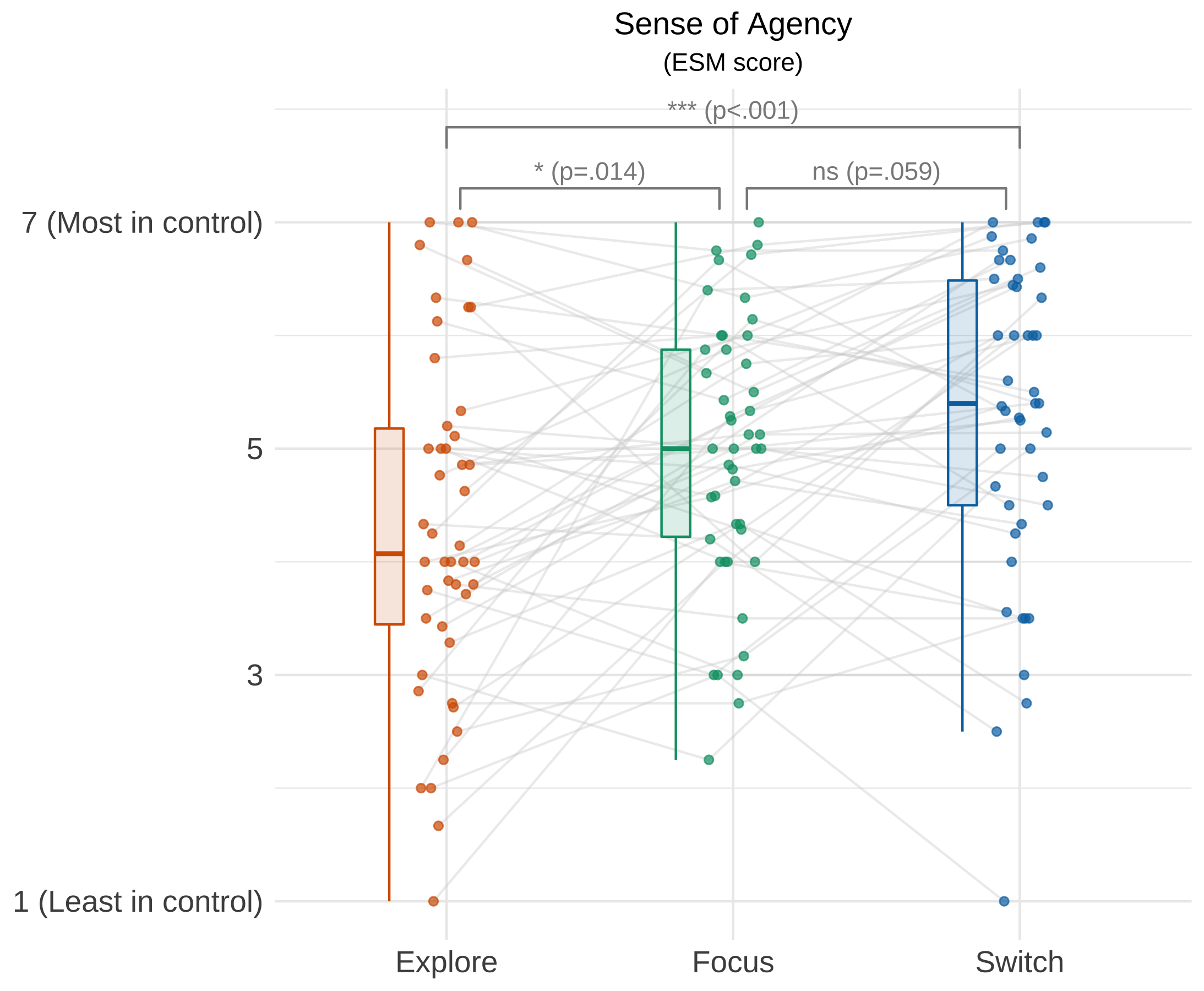}
    \caption{Experience sampling ratings for sense of agency were higher for Switch and Focus than Explore. Explore also had the greatest variation, with more participants feeling out of control as compared to the other versions. Note: Each dot represents the average of all of a single participant's ESM responses.}
    \label{fig9:senseofagency}
    \Description{A visualization showing the esm score ratings for sense of agency. In this visualization, Explore is lower than Focus with a median around 4, and Focus is lower than Switch with a median around 5. Switch has the highest median (around 5.4). 1 means least in control, and 7 means most in control }
\end{figure}

For sense of agency, our hypothesis was: Switch > Focus > Explore. The median of the participant means was 5.4 for Switch, 5 for Focus, and 4.1 for Explore (Figure \ref{fig9:senseofagency}). The participant means for control were statistically significantly different across conditions using the Friedman test of differences among repeated measures, $\displaystyle{\chi^2 (2) =15.3,\ p<0.001}$. A pairwise Wilcoxon signed rank test with a Holm-Bonferroni adjustment for multiple comparisons revealed statistically significant differences in participant mean score between Explore and Switch $\displaystyle{(Z=4.04,\ p<0.001)}$ and Explore and Focus $\displaystyle{(Z=2.71,\ p=0.14)}$, but not between Focus and Switch $\displaystyle{(Z=1.9,\ 0=0.059)}$. All subsequent ESM analyses use these same tests.\\

Because we had based the design of the SwitchTube app on features that participants reported as having an influence on their sense of agency, these hypotheses served as a manipulation check that was largely confirmed. The features in Focus did indeed increase sense of agency and the features in Explore decreased it. \\

The inability to customize recommendations led participants to feel like their attention had been led astray. One participant described not being able to find the Spanish-language music videos he was looking for in Explore: 
\begin{quote}
    I felt like I had no control, because it was only the videos that were on there and the search was not working very well at all. So I was limited to what somebody else chose for me. (P7)
\end{quote}
This sentiment also aligns with the findings of our previous work, where participants also found the current customization options in the YouTube mobile app (e.g., marking “Not Interested” on recommendations) to be wholly inadequate in terms of providing the level of control they desired \cite{27_Lukoff2021-gm}. \\

Switch was rated higher than Focus for sense of agency, but not quite significantly so $(p<0.059)$. Within Switch, the Focus Mode was generally preferred, but there were times when participants appreciated the option of toggling to Explore. In line with \textbf{Use Case 3} (between-session switching to address situational needs), one participant described:
\begin{quote}
    Oddly enough, my family collects different versions of the Monopoly board game just for fun… Since I was looking for videos [of new versions] late at night with my family, I just needed something really easy to do. Trying to fight through the search bar at that point was just too much effort. Just let me scroll through and find something interesting and we'll call it good. (P27)
\end{quote}
In other words, on occasion participants would `satisfice' by accepting recommendations that were suboptimal but low effort, rather than “fighting” the interface for the perfect video. While having the option of toggling on recommendations sometimes supported sense of agency, an inability to turn them off at all undermined sense of agency (as is the case in the Explore version and the current YouTube mobile app).

\subsubsection{\textbf{H2: Satisfaction}}
\begin{figure}[H]
    \centering
    \includegraphics[scale=0.10]{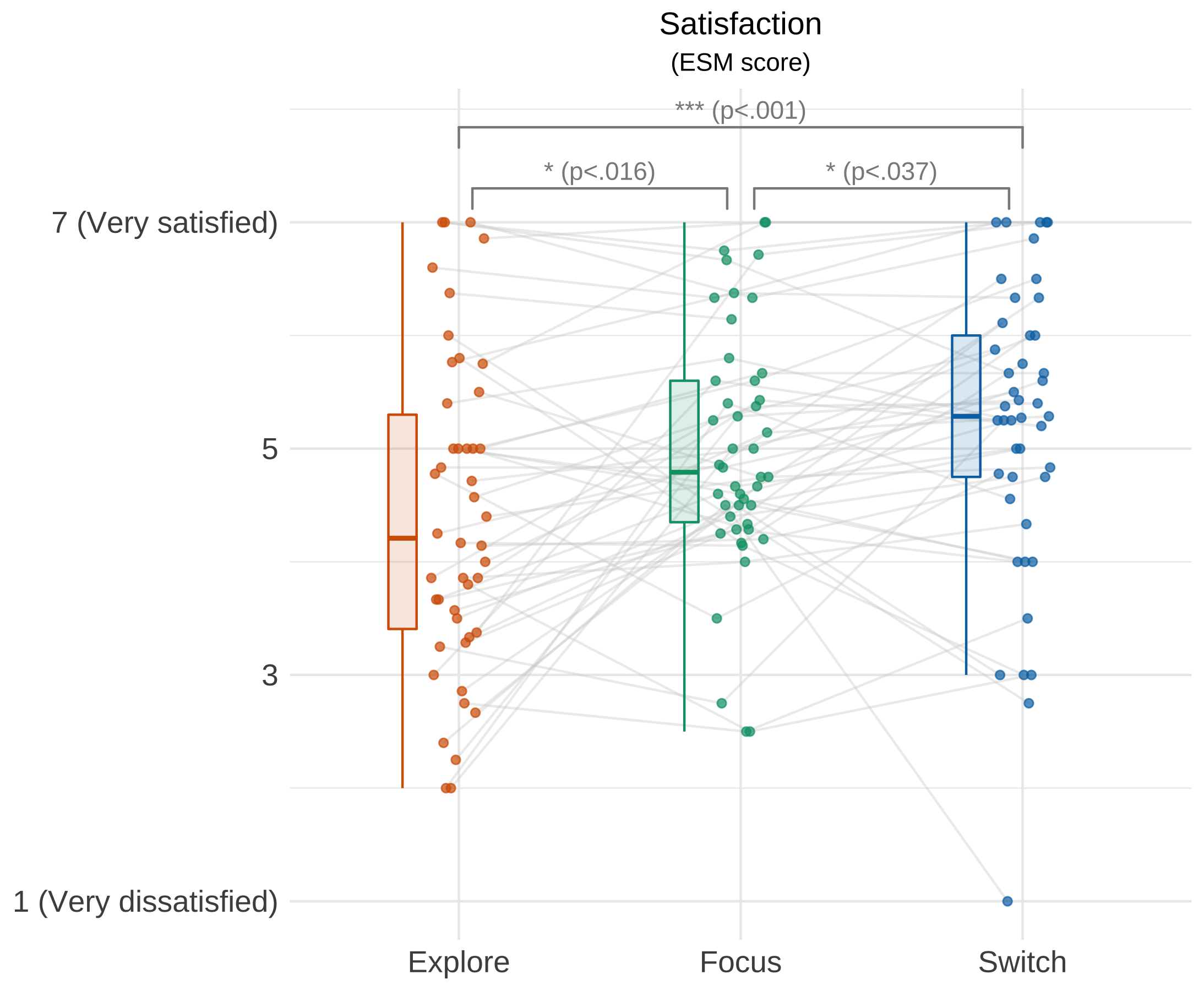}
    \caption{Experience sampling ratings for satisfaction were higher for Switch than for Focus and higher for Focus than for Explore. Again, Explore also had the greatest variation.}
    \label{fig10:satisfaction}
    \Description{A visualization showing the esm score ratings for satisfaction. In this visualization, Explore is lower than Focus with a median around 4.2, and Focus is lower than Switch with a median around 4.8. Switch has the highest median (around 5.4). 1 means very dissatisfied, and 7 means very satisfied.}
\end{figure}

For satisfaction, our hypothesis was: Switch > Explore > Focus. The median of mean scores per participant was $5.3$ for Switch, $4.8$ for Focus, and $4.2$ for Explore (Figure \ref{fig10:satisfaction}). Participant means for satisfaction were statistically significantly different across conditions: $\chi^2(2)=14.78, p<.001$. There were statistically significant differences in participant mean score between Explore and Switch $(Z=3.63, p<.001)$, Explore and Focus $(Z=2.66, p=.016)$, and Focus and Switch $(Z=2.1, p=.037)$.\\

Switch again rated highest, indicating that its mix of features supported short-term user satisfaction. However, contrary to our expectations, Focus was actually rated higher than Explore. Whereas we had expected that recommendations would contribute to short-term satisfaction, participants reported otherwise.

In interviews, participants appreciated the ability to switch between the interfaces to meet different needs. In line with \textbf{Use Case 2}, this satisfied curiosity to check out different content across the two versions:
\begin{quote}
    I switched back to Focus Mode, searched for Vivaldi, found “The Four Seasons,” and played that. And then, I guess, four hours later I opened it back up and was like, ‘Let's do something else,’ switched to Explore Mode and watched two random videos on vending machines. (P3)
\end{quote}
One might guess that the classical music listener is unlikely to watch videos about vending machines, but our participants did clearly enjoy exploring at times. Similarly, another user appreciated different search results for different occasions: \textit{“Both options are cool because one is relevant to the search terms and the other one is viral videos so I think both are effective”} (P9). The option to switch provided greater satisfaction.\\

Linking satisfaction back to sense of agency, many participants liked having the options that Switch afforded, even if they did not exercise them: \textit{“Switch did satisfy my needs more just because I had the option to switch between the different modes... Even if you don't use it, having the option is always good”} (P23). In other words, just having the feeling of being in control felt good. In cases like these, our research team found it difficult to disentangle whether a quote best described a participant’s sense of agency, satisfaction, or goal alignment. They were all related, which is also seen in the experience sampling ratings which follow a consistent preference order across all three of our measures: Switch >= Focus > Explore. 

\subsubsection{\textbf{H3: Goal Alignment}}
\begin{figure}[H]
    \centering
    \includegraphics[scale=0.10]{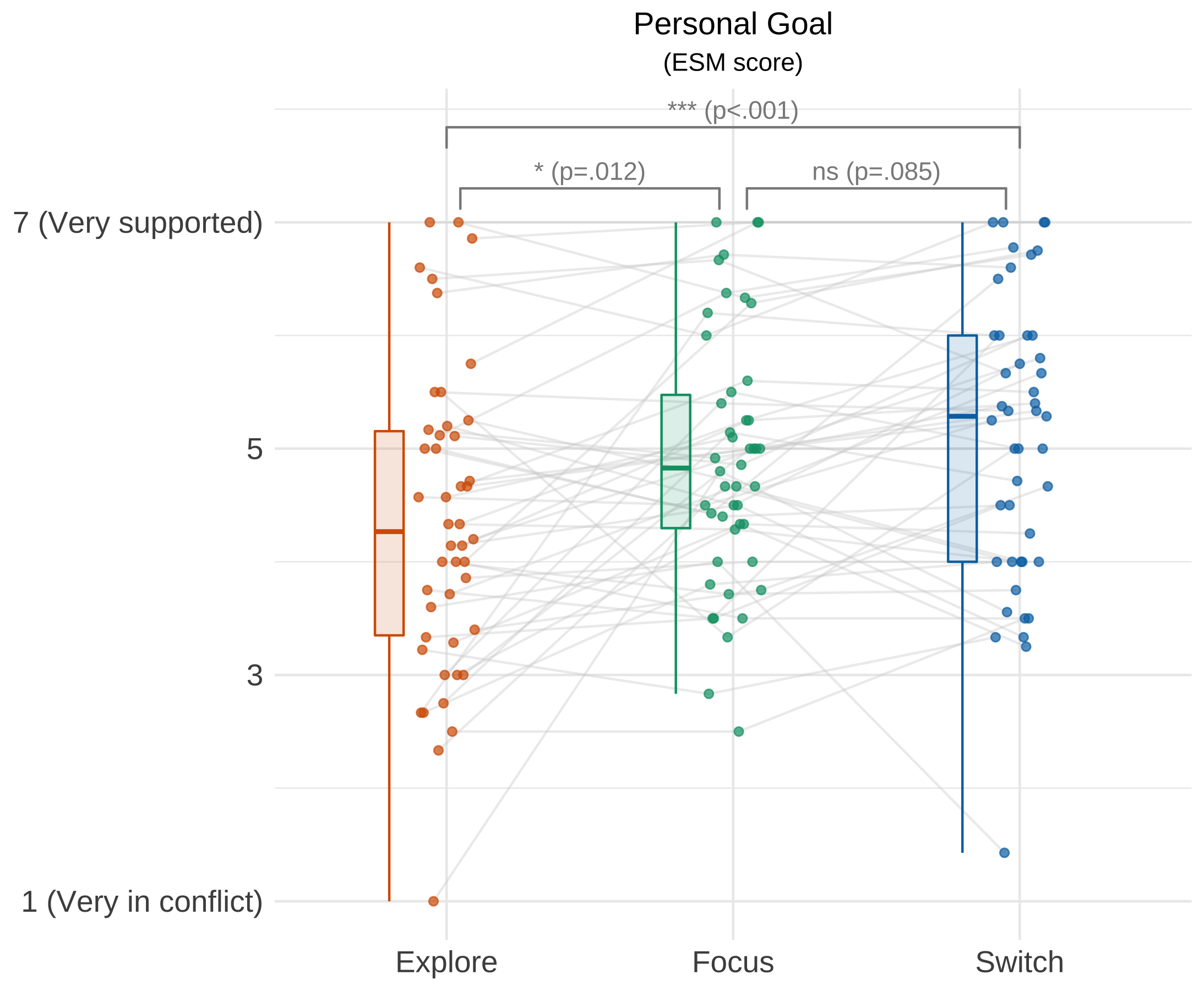}
    \caption{Experience sampling ratings for personal goal alignment were higher for Switch and Focus than Explore.}
    \label{fig11:personalgoal}
    \Description{A visualization showing the esm score ratings for personal goal. In this visualization, Explore is lower than Focus with a median around 4.3, and Focus is lower than Switch with a median around 4.8. Switch has the highest median (around 5.2). 1 means very in conflict,  and 7 means very supported.}
\end{figure}

For H3, our expectation was: Switch > Focus > Explore. The median of mean scores per participant was $5.3$ for Switch, $4.8$ for Focus, and $4.3$ for Explore (Figure \ref{fig11:personalgoal}). Participant means for goals were statistically significantly different across conditions: $\chi^2(2)=$ 13.32, $p<.001$. There were statistically significant differences in participant mean score between Explore and Switch $(Z=3.96, p<.001)$ and Explore and Focus $(Z=2.74, p=.012)$, but not Focus and Switch $(Z=1.73, p=.085)$. 

The limited number of features in Focus supported goal alignment. With just a search bar and video feed feature on the home page, the minimal design made users feel \textit{``less likely to get distracted''} by recommendations and related videos. Instead, users felt it was easier to focus on the videos they intended to watch and spend less time browsing the app. For example, in the initial survey, P33 expressed their high-level goal as the following: \textit{``I want to be able to use this app less than I am used to.''} When interviewed, they explained that Focus best aligned with this goal:
\begin{quote}
    I liked how in Focus it was just the video and nothing else. There was no comments to read, no suggested videos on the bottom to scroll through. So, I feel like I was very focused on that one video that I was watching.
\end{quote}
Without the distraction of comments or video recommendations that they felt \textit{``forced''} to watch,  users were able to quickly accomplish what they were set to do in SwitchTube, close the app, and do something else. Knowing that Focus does not show recommendations, participants used the app when they had a specific goal in mind as described by P6 who wanted to reduce the quantity of time they spent browsing: \textit{“It was more, set a goal, find it, and complete that goal instead of just scrolling through randomly looking for something.”}\\

In Switch, participants commonly described experiences that aligned with \textbf{Use Case 4}, within-session switching, where the participant would start focused and then explore from there. P19 described how in Switch they could say:
\begin{quote}
    ‘I don't want to watch that. I'm only going to go in these directions.’ So even though I spent less time, the majority of it felt very, very much like I wanted to watch the initial video that I watched, and I felt good about doing that.
\end{quote}
Using an initial video to direct their exploration helped participants stick to their original goal for use and feel satisfied about it.

Focus better supported participants even when they spent the same amount of time in it as Explore because the quality of the content that they watched better aligned with their goals. The same participant described:
\begin{quote}
    [With Focus] it's easier to feel satiated: ‘Okay, yeah, that was cool. I'm glad I watched that. I don't feel like I necessarily wasted my time’… And I think it's funny to see that I spent more time in the Focus, but the quality of it felt better. (P19)
\end{quote}
In contrast to Explore, participants said Focus did not induce the same feelings of guilt about \textit{``wasting time.''} 

\section{Discussion}
Overall, in our survey we found that about half of the participants in our sample of general YouTube users were either actively or imminently trying to change something about their YouTube use. Reducing the quantity of time spent on YouTube was a common goal, but so was changing the quality of the content, for example by increasing consumption of educational videos or decreasing that of recommendations. \\

In our field experiment, Switch better supported sense of agency, satisfaction, and goal alignment than Explore alone, and also offered greater satisfaction than Focus alone. Focus alone was also rated significantly higher than Explore on all three of these measures. Unexpectedly, this was true despite the fact that time spent in Focus (median: 145 minutes) was actually somewhat higher than in Explore (median: 117). In terms of switching behavior, participants engaged in both between-session switching to stick to an original intention for that session of use (\textbf{Use Case 3}) and within-session switching to start focused and then explore from there (\textbf{Use Case 4}). When Explore was the only option it was rated relatively low in terms of user experience, but the high ratings for Switch and the prevalence of Use Cases 3 and 4 show that there were situations where users did find it beneficial. We discuss these findings and their design implications.

\subsection{Quality of Time, Not Just Quantity}
For the screen time tools that come pre-installed on every iOS and Android smartphone (Apple Screen Time and Google’s Digital Wellbeing) and in the vast majority of third-party digital wellbeing apps and extensions \cite{30_Lyngs2019-xr}, the goal is to reduce time spent on the phone or in certain apps. However, our survey results add to the growing chorus of digital wellbeing researchers calling for people to move beyond just screen time as the metric of success \cite{7_Cecchinato2019-td,11_Devito2019-qi,28_Lukoff2019-wc}. Specifically, in the case of YouTube, we find that decreasing quantity of time accounted for only 37\% of the goals that survey participants had. A larger category of goals was shifting the quality of consumption (51\% in total), either by increasing a certain type of content (e.g., educational, new, or meaningful videos) or decreasing another (e.g., meaningless, sponsored, or recommended videos). In other words, screen time did matter for participants, but it was a less common goal than changing the quality of time spent on YouTube, which is largely unsupported by current screen time tools.\\

Our results also contribute empirical evidence that designers can support concerned users and improve user experience without necessarily reducing screen time. To our surprise, SwitchTube study participants actually spent more time in Focus (median: 145 minutes) than in Explore (median: 117), and although many of them had a goal of reducing their time spent on YouTube, they still rated it significantly higher than Explore in terms of sense of agency, satisfaction, and goal alignment. Participants said that even if they spent the same amount of time in Focus as in Explore, that time was usually better spent. Similar results have recently been found for a version of Twitter that was redesigned to support a greater sense of agency \cite{49_Zhang2022-va}. The term ``screen time'' has provided a common vocabulary for many users to voice their concerns with digital device use, but it has also led to an overemphasis on \textit{time} as the singular metric of success and limited the solutions imagined by design researchers and practitioners. \\

One key consideration in analyzing our experiment is that homepage recommendations were not personalized in SwitchTube as they are in the YouTube app. Since recommendations (both personalized and non-personalized) account for about 70\% of watch time on YouTube \cite{Solsman2018-wn}, this means that the Explore Version may be difficult to compare against the current YouTube app. Adding personalized recommendations to SwitchTube would likely have increased the number of recommended videos watched and the amount of time spent in the Explore Version, but not in the Focus Version in which homepage recommendations were turned off by default. We expect that personalized recommendations would have reduced irrelevant recommendations but also increased temptation, perhaps leading Explore to be rated higher on satisfaction but lower on goal alignment—as we had originally hypothesized. Future work could investigate how personalized recommendations influence time spent and user experience ratings from the perspective of digital wellbeing. %

\subsection{Adaptable Commitment Interfaces for Digital Wellbeing}
The term “adaptable commitment interface” sounds like a paradox. How can one “commit” if one can easily switch away from that commitment at any time without any cost? An ACI lacks the core mechanism behind commitment devices: that the present self imposes a meaningful cost (in terms of time, money, or effort) upon the future self for abandoning their original goal. Unlike the concept of the goal reminders that are also used in the digital wellbeing space \cite{18_Hiniker2016-cs,31_Lyngs2020-jp}, an ACI does not necessarily provide informational support. Yet the ACI we designed (the Switch Version of SwitchTube) received higher participant ratings for sense of agency, satisfaction, and goal alignment than an interface that lacked commitments (the Explore Version of SwitchTube). The ACI also outperformed a non-adaptable commitment interface (the Focus Version of SwitchTube) in terms of participant ratings for satisfaction. \\

We believe that Switch worked as an ACI because it avoided triggers for distraction in the first place. This aligns with the emerging consensus in psychological research that people who are effective self-regulators cannot necessarily use willpower to resist temptations once encountered, but are instead skilled at avoiding them altogether \cite{12_Duckworth2020-kn, 13_Duckworth2016-mq}. For digital wellbeing designers, this suggests that rather than waiting to intervene “just-in-time” by reacting to tempting content with a barrier or goal reminder, they should consider how to prevent that content from appearing in the first place. Of course, this could also be implemented by just entirely blocking or limiting access to an app with distracting content as is currently done in many screen time tools (a commitment interface), but the cost is that it can also indiscriminately block access to desired features and provoke psychological reactance in response to the restrictions it imposes on user autonomy \cite{26_Lukoff2022-xs,29_Lyngs2022-ug}. An advantage of ACIs is that they impose no such autonomy restriction: in Switch, if the user wanted to toggle over to Explore Mode, they were free to do so. Participants reported a high sense of agency in the Switch Version.\\

One clear design implication is that users would benefit from a ``Focus Mode'' even within an app such as YouTube that is commonly regarded as a service for entertainment. And of course, YouTube does provide entertainment, but users often have goals of using it for other purposes as well---for example, education both formal and informal, as in the SwitchTube participants who described watching videos to learn Spanish and to fix a flat tire. Rather than offering Focus Mode only in “productivity” tools like Microsoft Word (where there are far fewer distractions to begin with), there is a strong case for adding it to services that people rely upon for many different use cases, such as YouTube, Twitter, and Facebook. \\

Unfortunately our finding that a Focus Mode would benefit the user experience does not mean that social media companies will necessarily be keen adopt it. In fact, the design of social media seems to be trending in the opposite direction: short videos that play on continuous autoplay have powered the rise of TikTok and are now being emulated by YouTube, Facebook, and Instagram \cite{McCluskey2022-dq}. As a matter of practicality, designers needing to convince other interested parties that a Focus Mode will not necessarily harm the business interests of advertising-supported services might cite our result that the Focus Version did not reduce time spent in SwitchTube.\footnote{However, as discussed earlier, this result may have been different if the Explore Version had of included \textit{personalized} video recommendations.} \\

As a matter of principle though, designs that support user agency and wellbeing should be encouraged \textit{even if} they go against the business interests (e.g., reduce time spent on a service). As such, our findings may also be of interest to policymakers and regulators who are tasked with crafting and implementing incentives for online platforms. For example, the Digital Services Act that was approved by the European Council in October 2022 calls for an end to dark patterns that \textit{"distort and impair user autonomy, decision-making and choice"} \cite[p.57]{European_Council2022-eg}. Translating such an ambitious and sweeping directive into action will require further studies in the vein of SwitchTube that attempt to enact and assess abstract values such as `user sense of agency' in digital interfaces.

\subsection{From Adaptable to Adaptive Commitment Interfaces}
SwitchTube is an \textit{adaptable} commitment interface, but it also holds implications for \textit{adaptive} commitment interfaces. In Switch, the user can manually toggle between two modes, one that lets the system take control (Explore Mode) and the other that helps hold the user to their intention (Focus Mode). Yet this manual labor does place a certain burden upon the user and might instead be automatically performed by the system itself, i.e., an adaptive commitment interface. For example, if YouTube has high confidence that the user is visiting with a specific intention, it could present a search-only interface and hide all recommendations. Conversely, if YouTube expects that the user is visiting without an intention in mind, it could take control as in its Leanback mode for “effortless viewing” that autoplays a never-ending stream of recommendations \cite{15_Google_undated-yh}. %

Already, machine-learning systems can predict the specificity of a user's intent for visiting a site with some accuracy \cite{8_Cheng2017-ge,20_Hiniker2016-ef}, opening the door for interfaces that adapt the level of sense of agency to meet a user’s needs. Yet given that these predictions do still have a substantial error rate at present and erroneous predictions could provoke user reactance, what might a more reliable first step towards such an adaptive commitment interface look like? \\

\textit{\textbf{Use Case 4}}, within-session switching, suggests one such direction. Participants started in Focus Mode to search and then switched to Explore Mode after they had performed their search to branch out from there. Because search queries are a strong signal of intent, this could be an ideal opportunity for the system to make a well-informed prediction of the user's needs and adapt the interface accordingly. For example, such a system would always start in the search-first Focus Mode. Searches such as ``flat tire'' that suggest a specific intention would keep the user in Focus Mode whereas a query such as ``standup comedy'' that suggests an openness to simply being entertained would automatically switch them to Explore Mode. Such a search-informed adaptive commitment interface could offer semi-automatic support for when a user has a focused goal, but still afford the flexibility to explore. In other words, the user provides the general direction of their intention and the system supports them to optimize for that intent, adapting into the interface for either focus or exploration.

\subsection{Limitations and Future Work}
A limitation of our work is that we compared the time spent in each of the three versions of SwitchTube, but did not measure what, if any, spillover effect there was on time spent on YouTube itself. Prior work has found that productivity interventions on one platform do not redistribute time to other apps or platforms \cite{Kovacs2019-pe}. In the case of SwitchTube though, it is possible that because Explore is the most similar version to the existing interface of YouTube it was seen as a poor substitute, whereas Focus and Switch offered something different, so were used more often as a substitute for YouTube use. \\

Our findings also highlight an important need for research on how to help users \textit{shift} their use in desired directions, not just how to reduce it. Again due to limitations of the YouTube Data API, we had to make due with kludgy workarounds (e.g., appending ``viral'' to the search query) to change the quality of the content in our study app. Future work in digital wellbeing might devise clever ways to manipulate the inner workings of recommender systems and search algorithms that drive so much of social media use today. To inform the design of such algorithms, one starting point could be to analyze the content of the dataset of 2063 video play events in SwitchTube that we make available via the Open Science Framework: \href{https://osf.io/z735n/}{\color{blue}{https://osf.io/z735n}}

\section{Conclusion}
Our work steps beyond the current paradigm of screen time tools and demonstrates how a digital wellbeing design might also support users in shifting the \textit{quality} of their consumption. We designed and developed SwitchTube, an Android app with three different versions for watching YouTube videos, and deployed it in a field experiment with 46 US YouTube users over three weeks. Focus was rated higher than Explore on user experience metrics, but surprisingly it did not reduce the amount of time that participants spent watching videos. Yet the Switch Version, which provided the ability to toggle between both Explore and Focus Mode, was rated the highest of all. In other words, the Switch Version proved effective as an adaptable commitment interface (ACI) that provided the user with the flexibility to switch, but still helped them to follow their initial intention, providing the best of both worlds. Designers interested in digital wellbeing should consider implementing ACIs that support the user to explore when they want to explore and focus when they want to focus. 

\begin{acks}
This work was funded in part by National Science Foundation award \#1849955. Ulrik Lyngs is supported by the Carlsberg Foundation [grant number CF20-0678]. We thank Anqi Cao, James Choi, Kaiyue Fan, Linda Lai, Jiangzhenjun (Vera) Liao, Lynn K. Nguyen, and Xuecong (Esme) Xu for helping to conceptualize, design, and test the SwitchTube app.
\end{acks}

\bibliographystyle{ACM-Reference-Format}
\bibliography{references}

\end{document}